\documentclass[a4paper,hidelinks]{article}\usepackage[]{graphicx}\usepackage{xcolor}
\makeatletter
\def\maxwidth{ %
  \ifdim\Gin@nat@width>\linewidth
    \linewidth
  \else
    \Gin@nat@width
  \fi
}
\makeatother

\definecolor{fgcolor}{rgb}{0.345, 0.345, 0.345}
\newcommand{\hlopt}[1]{\textcolor[rgb]{0,0,0}{#1}}%
\newcommand{\hlstd}[1]{\textcolor[rgb]{0.345,0.345,0.345}{#1}}%
\newcommand{\hlkwd}[1]{\textcolor[rgb]{0.737,0.353,0.396}{\textbf{#1}}}%

\usepackage{framed}
\makeatletter
\newenvironment{kframe}{%
 \def\at@end@of@kframe{}%
 \ifinner\ifhmode%
  \def\at@end@of@kframe{\end{minipage}}%
  \begin{minipage}{\columnwidth}%
 \fi\fi%
 \def\FrameCommand##1{\hskip\@totalleftmargin \hskip-\fboxsep
 \colorbox{shadecolor}{##1}\hskip-\fboxsep
     \hskip-\linewidth \hskip-\@totalleftmargin \hskip\columnwidth}%
 \MakeFramed {\advance\hsize-\width
   \@totalleftmargin\z@ \linewidth\hsize
   \@setminipage}}%
 {\par\unskip\endMakeFramed%
 \at@end@of@kframe}
\makeatother

\definecolor{shadecolor}{rgb}{.97, .97, .97}
\definecolor{messagecolor}{rgb}{0, 0, 0}
\definecolor{warningcolor}{rgb}{1, 0, 1}
\definecolor{errorcolor}{rgb}{1, 0, 0}
\newenvironment{knitrout}{}{} % an empty environment to be redefined in TeX

\usepackage{alltt}
\usepackage{bm,amsmath,amssymb,amsthm,amsfonts,graphics,graphicx,epsfig,
rotating,caption,subcaption,natbib,appendix,titlesec,multicol,hyperref,verbatim,
bbm,algorithm,algpseudocode,pgfplotstable,threeparttable,booktabs,mathtools,
dsfont,parskip,xr,enumerate,multirow,tikz,arydshln}
\usetikzlibrary{matrix, fit, backgrounds}
\usepackage{titling}

\newcommand{\argmax}{\text{argmax} \,}
\newcommand{\expit}{\text{expit}}

\newcommand{\E}{\mathbb{E}}

\newcommand{\tr}{^{\text{T}}}
\newcommand{\diag}{\text{diag}}

\newcommand{\appropto}{\mathrel{\vcenter{
  \offinterlineskip\halign{\hfil$##$\cr
    \propto\cr\noalign{\kern2pt}\sim\cr\noalign{\kern-2pt}}}}}

\makeatletter
\newcommand*{\defeq}{\mathrel{\rlap{%
			\raisebox{0.3ex}{$\m@th\cdot$}}%
		\raisebox{-0.3ex}{$\m@th\cdot$}}%
	=}
\makeatother

\renewcommand{\(}{\left(}
\renewcommand{\)}{\right)}
\renewcommand{\[}{\left[}
\renewcommand{\]}{\right]}

\makeatletter
\newcommand{\vast}{\bBigg@{3}}
\newcommand{\vastt}{\bBigg@{4}}
\newcommand{\Vast}{\bBigg@{5}}
\makeatother

\pgfplotsset{compat=1.16}
\graphicspath{{../figs/}}
\title{Semi-supervised empirical Bayes group-regularized factor regression}
\date{\today}
\author{Magnus M. M\"unch$^{1,2}$\protect\footnote{Correspondence to: 
\href{mailto:m.munch@amsterdamumc.nl}{m.munch@amsterdamumc.nl}}, Mark A. van de 
Wiel$^{1,3}$, \\ Aad W. van der Vaart$^{2}$, and Carel F.W. Peeters$^{1,4}$}
\IfFileExists{upquote.sty}{\usepackage{upquote}}{}
\begin{document}

	\maketitle
	
	\noindent
	1. Department of Epidemiology \& Data Science, Amsterdam UMC, VU University, 
	PO Box 7057, 1007 MB Amsterdam, The Netherlands \\
	2. Mathematical Institute, Leiden University, Leiden, The Netherlands \\
	3. MRC Biostatistics Unit, Cambridge Institute of Public Health, Cambridge,
	United Kingdom \\
	4. Division of Mathematical \& Statistical Methods - Biometris, 
	Wageningen University \& Research, Wageningen, The Netherlands
	
	\begin{abstract}
		{
		The features in high dimensional biomedical prediction problems are often
		well described with lower dimensional manifolds. An example is
		genes that are organised in smaller functional networks. The outcome can 
		then be
		described with the factor regression model. A
		benefit of the factor model is that is allows for straightforward inclusion
		of unlabeled observations
		in the estimation of the model, i.e., semi-supervised learning.
		In addition, the high dimensional features in biomedical prediction problems
		are often well characterised. Examples are genes, for which annotation is
		available, and metabolites with $p$-values from a previous study available.
		In this paper, the extra information on the features is included in the 
		prior model for the features. 
		The extra information is weighted and included in the estimation
		through empirical Bayes, with Variational approximations to speed up the
		computation.
		The method is demonstrated in simulations and two applications.
		One application considers influenza vaccine 
		efficacy prediction based on microarray data. The second application
		predictions oral cancer metastatsis from RNAseq data.
		}
	\end{abstract}
	
	
	\noindent\textbf{Software available from}: 
	\url{https://github.com/magnusmunch/bayesfactanal}
	
	\section{Introduction}
	In modern biomedical research, there is an interest in 
	prediction models based on large sets of omics features. Common 
	outcomes are, for example, categorical disease status, 
	time-to-event, or continuous anthropomorphic measures.
	
	In many omics studies, the number of omics features considered is large
	and may run in the tens of thousands (in, e.g., genomics). At the same time,
	the number of samples is generally low, commonly due to 
	high measurement costs, logistics, or the availability of subjects. 
	The high-dimensionality of 
	data (i.e., $p > n$) complicates model estimation.
	On the other hand, extra unlabeled omics data is often available.
	Here, `unlabeled' refers to data for which the predictor features are
	available, but not the study response/outcome.
	Unlabeled data may, for example, come from online repositories or 
	previous studies with the same set of features, but a different response.
	Inclusion of unlabeled data in prediction problems, termed semi-supervised
	learning in the machine learning community, has received plenty of attention
	\cite[see][for an introduction]{zhu_introduction_2009}.
	
	Several authors have argued that the high-dimensional feature space in omics 
	data arises from noisy observations on a lower dimensional latent space. 
	\cite{bernardo_bayesian_2003} show that gene 
	expression data from breast cancer patients are indeed well described with a
	lower dimensional (linear) latent space.
	Moreover, \cite{carvalho_high-dimensional_2008} improve prediction of 
	mutant p53 gene versus wild type in breast cancer patients with the lower
	dimensional structure of the gene expression data. 
	\cite{bernardo_bayesian_2003} and \cite{carvalho_high-dimensional_2008} use
	a Bayesian linear factor (regression) model approach to describe the latent 
	space. \cite{mes_outcome_2020} is an example of a frequentist
	latent space approach (technically a hybrid between Bayes and frequentist) to
	prediction from radiomics features.
	
	Inclusion of unlabeled data into the estimation of
	linear factor models may benefit estimation 
	\cite[]{liu_maximum_1998,banbura_maximum_2014}. Here, we extend the estimation
	of a Bayesian factor regression model to include unlabeled data to improve 
	prediction from a high-dimensional feature space. We treat the
	unlabeled data as data with a missing response and consider the full 
	likelihood approach, together with a Bayesian prior. 
	
	In addition, extra information on the features is often available. The extra
	information, termed co-data, may be a partitioning of the features, such as 
	pathway membership of the genes, or continuous information, such as $p$-values
	from a previous study. Recently, several methods have been introduced that
	use the co-data to improve prediction 
	\cite[see, e.g.,][]{van_nee_flexible_2020,munch_adaptive_2019,
	te_beest_improved_2017,van_de_wiel_better_2016}. 
	
	In the current paper, we apply the co-data approach
	\cite[more specifically, a group-adaptive empirical Bayes approach akin to 
	that in][]{munch_adaptive_2019}, together with the
	inclusion of the unlabeled data, to the
	Bayesian factor regression model. We present an extension 
	of the method to a mixed mode factor analysis, the outcome is
	binary instead of continuous. Simulations show that the approach is
	competitive or even outperforms classical approaches in some settings. 
	Applications to influenza vaccine efficacy 
	prediction and oral cancer lymph node metastasis prediction show that the
	approach has the potential to enhance predictive performance compared to
	existing methods.
	
	The rest of the paper is organised as follows: Sections \ref{sec:model} 
	and \ref{sec:estimation} describe the model and its estimation in detail. 
	The approach is demonstrated in a simulated setting in Section
	\ref{sec:simulations} and two
	real data settings in Section \ref{sec:applications}. We conclude with
	a short discussion on the pros and cons of the method in Section 
	\ref{sec:discussion}.
	
	\section{Model}\label{sec:model}
	\subsection{Observational model}
	We assume our observed $p$-dimensional feature vectors $\mathbf{x}_i$ and 
	outcomes $y_i$, $i=1, \dots n$, are standardised, such that 
	$\forall j: \sum_{i=1}^n x_{ij}=0$, 
	$\forall j: \sum_{i=1}^n x^2_{ij}=n$,
	$\sum_{i=1}^n y_{i}=0$, $\sum_{i=1}^n y_{i}^2=n$, and 
	follow the factor regression model \cite[]{liang_use_2007}:
	\begin{subequations}\label{eq:model}
	  \begin{align}
	    y | \bm{\lambda} & \sim \mathcal{N}(\bm{\beta}^{\text{T}} \bm{\lambda},
	      \sigma^2), \label{eq:linearoutcome}\\
	    \mathbf{x} | \bm{\lambda} & \sim \mathcal{N}_p(\mathbf{B}^{\text{T}} 
	      \bm{\lambda}, \bm{\Psi}) \label{eq:linearfeatures}\\
	    \bm{\lambda} & \sim \mathcal{N}_d(\mathbf{0}, \mathbf{I}_d),
	      \label{eq:linearfactors}
	  \end{align}
	\end{subequations}
	where $\bm{\lambda}$ consists of the latent factors, 
	$\bm{\Psi}=\text{diag}(\psi_j)$, $j=1\dots, p$, are the uniquenesses,
	$\sigma^2$ is the error variance, and $\mathbf{B}$ and $\bm{\beta}$ are 
	the factor loadings. The latent factor dimension $d$ is 
	assumed fixed and known. The factor model comes with some issues
	(namely, rotational invariance and factor indeterminancy). These do
	not play a major role in prediction problems, so we do not address them here.
	SM Section \ref{sec:modelunidentifiability} provides some pointers into
	these issues. Model (\ref{eq:model}) implies a joint multivariate Gaussian 
	distribution for $\begin{bmatrix} \mathbf{x}^{\text{T}} & y 
	\end{bmatrix}^{\text{T}}$ (unconditional on $\bm{\lambda}$), 
	so a prediction $\tilde{y}$ from observed features
	$\tilde{\mathbf{x}}$ is obtained by taking the expectation of the
	conditional distribution of $\tilde{y}$ given $\tilde{\mathbf{x}}$ from
	model (\ref{eq:model}): 
	\begin{equation}\label{eq:prediction}
	  \mathbb{E}(\tilde{y}|\tilde{\mathbf{x}})=
	  \tilde{\mathbf{x}}^{\text{T}}
	  (\mathbf{B}^{\text{T}}\mathbf{B} + \bm{\Psi})^{-1}\mathbf{B}^{\text{T}}
	  \bm{\beta}=:\tilde{\mathbf{x}}^{\text{T}}\tilde{\bm{\beta}}.
	\end{equation}
	
	In the following, it is convenient to write $\bar{p}=p +1$,
	$\bar{\mathbf{x}} = 
	\begin{bmatrix}\mathbf{x}^{\text{T}} & y \end{bmatrix}^{\text{T}}$,
	$\bar{\mathbf{B}}=\begin{bmatrix}\mathbf{B} & 
	\bm{\beta}\end{bmatrix}$,
	$$
	\bar{\bm{\Psi}}=\begin{bmatrix} 
	\bm{\Psi} & \mathbf{0}_{p \times 1} \\
	\mathbf{0}_{1 \times p} & \sigma^2,
	\end{bmatrix}
	$$ 
	and consider the equivalent form of (\ref{eq:model}):
	\begin{subequations}\label{eq:model2}
	  \begin{align}
	    \bar{\mathbf{x}} | \bm{\lambda} & \sim \mathcal{N}_{\bar{p}}( 
	      \bar{\mathbf{B}}^{\text{T}} \bm{\lambda}, \bar{\bm{\Psi}}) \\
	    \bm{\lambda} & \sim \mathcal{N}_d(\mathbf{0}, \mathbf{I}_d).
	  \end{align}
	\end{subequations}
	
	If the outcomes $y_i$ are of sums of $N_i$ disjoint binary events with
	the shared probability of success, the linear outcome model 
	(\ref{eq:linearoutcome}) is replaced with its logistic counterpart:
	\begin{equation}\label{eq:logisticoutcome}
	y | \bm{\lambda}, \bm{\beta}, \beta_0 \sim 
	\mathcal{B}\left(N, \expit(\beta_0 + \bm{\beta}^{\text{T}}
	\bm{\lambda})\right),
	\end{equation}
	where $\mathcal{B}\left(N, \pi\right)$ denotes the binomial distribution with
	number of trials $N$ and success probability $\pi$. Note that the logistic
	model includes an intercept $\beta_0$ to accommodate unbalanced data, 
	whereas the linear model simply considers standardised data.
	Feature and factor models (\ref{eq:linearfeatures}) and 
	(\ref{eq:linearfactors}), in combination with outcome model
	(\ref{eq:logisticoutcome}) result in a mixed-mode factor model, with Gaussian 
	and binomially distributed features and outcome, respectively.
	This mixed-mode extension is detailed in Section 
	\ref{sec:logistic} of the Supplementary Material (SM).
	
	\subsection{Bayesian prior model}
	In the Bayesian version of the model, the parameters $\theta := 
	\{ \bar{\mathbf{B}},\bar{\psi}_1,\dots,
	\bar{\psi}_{p+1}\}$ are endowed with conditionally conjugate prior 
	distributions:
	\begin{subequations}\label{eq:prior}
  	\begin{align}
  	  \bar{\mathbf{B}} | \bar{\psi}_1, \dots, \bar{\psi}_{\bar{p}} & \sim 
  	    \prod_{\bar{j}=1}^{\bar{p}} \mathcal{N}_{d}(\mathbf{0}_{d}, 
  	  \bar{\psi}_{\bar{j}} \gamma_{\bar{j}} \mathbf{I}_d), \\
  	  \bar{\psi}_1, \dots, \bar{\psi}_{\bar{p}} & \sim \prod_{\bar{j}=1}^{\bar{p}} 
  	    \Gamma^{-1}(\kappa_{\bar{j}}, \nu_{\bar{j}}),
  	\end{align}
  \end{subequations}
  where $\Gamma^{-1}(\kappa,\nu)$ denotes the inverse 
  Gamma distribution with shape $\kappa$ and scale $\nu$. 
  Note that index $\bar{j}$ and dimension $\bar{p}$  
  indicate the use of the equivalent model formulation (\ref{eq:model2}).
  In addition, we write $\bar{\psi}_{\bar{p}}:=\sigma^2$ for notational 
  convenience. Here, the variance of $\bar{\mathbf{b}}_j$ 
  (column $\bar{j}$ of $\bar{\mathbf{B}}$) scales with 
  error variance/uniquenesses
  $\bar{\psi}_{\bar{j}}$ as is common in Bayesian
  (univariate) linear models. This is mostly for computational reasons, but is
  often justified as a solution to scaling problems in multivariate 
  regression problems \cite[]{leday_gene_2017}. 
  
  In the Bayesian model a prediction $\tilde{y}$ from features 
  $\tilde{\mathbf{x}}$ is obtained by averaging over the posterior: 
	\begin{equation}\label{eq:bayesianprediction}
	  \mathbb{E}^*(\tilde{y}|\tilde{\mathbf{x}})=
	  \tilde{\mathbf{x}}^{\text{T}}
	  \mathbb{E}_{\bar{\mathbf{B}}, \bar{\bm{\Psi}} | \bar{\mathbf{x}}}
	  \[(\mathbf{B}^{\text{T}}\mathbf{B} + 
	  \bm{\Psi})^{-1} \mathbf{B}^{\text{T}}
	  \bm{\beta}\]=:\tilde{\mathbf{x}}^{\text{T}}\bm{\beta}^*.
	\end{equation}
	In practice, this expectation is hard to compute. 
	Here, we use a combination of variational Bayes for posterior computation and 
	Monte Carlo simulation for approximation of (\ref{eq:bayesianprediction}). 
	An alternative to Monte Carlo simulation is Taylor approximation, as explained 
	in Section \ref{sec:posteriorexpectation} of the SM.
	
	\subsection{Additional feature structure}
\begin{figure}[h!]
	  \centering
    \includegraphics[width=0.8\linewidth]{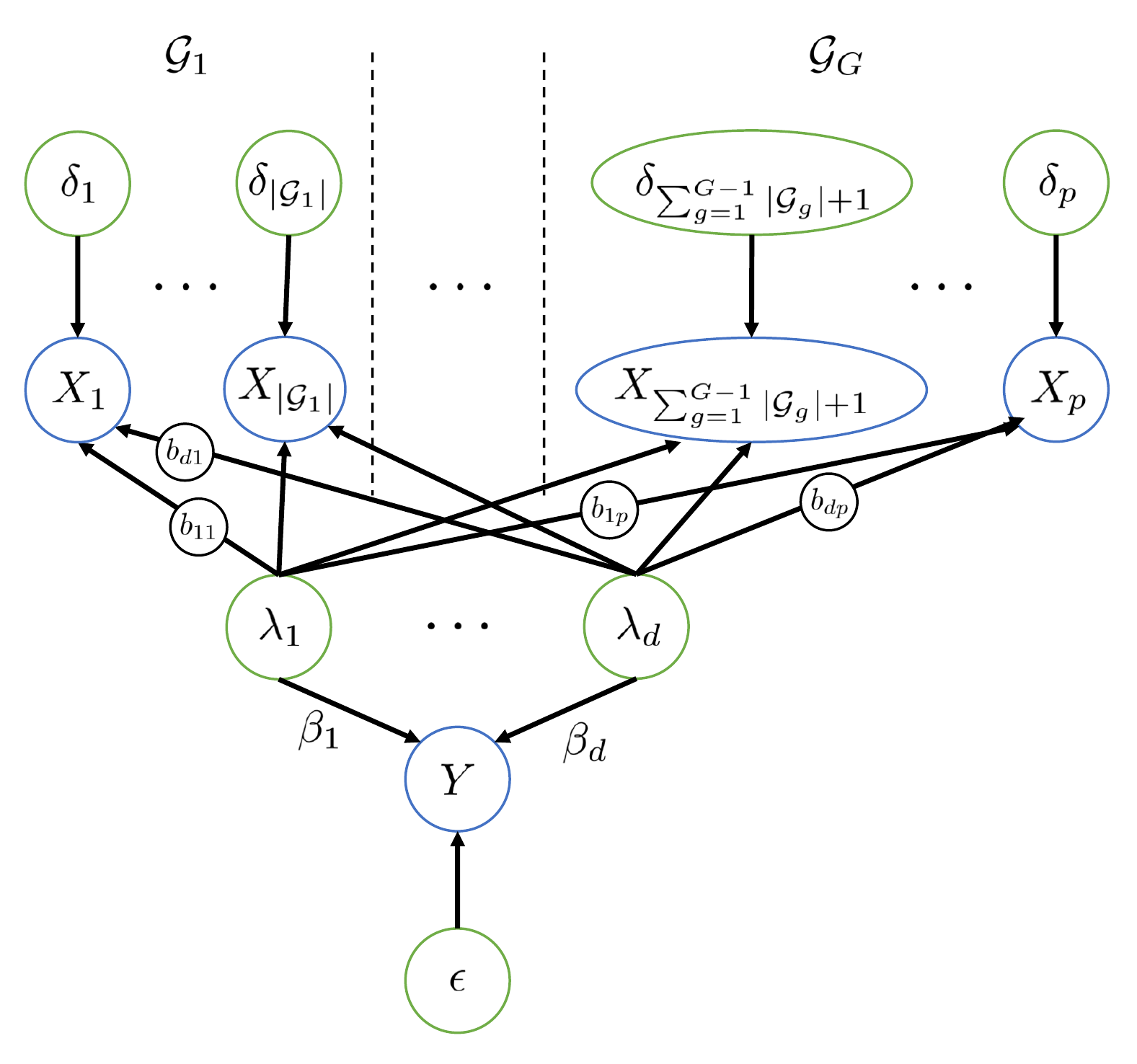}
    \caption{Model (\ref{eq:model}) with paritioned features as a Bayesian 
    network, where the vertical dotted lines denote a partitioning of features 
    $X_1, \dots, X_p$ into groups $g=1, \dots, G$. Green and blue circles denote
    latent and observed variables, respectively. Note that the $\delta_j$ and
    $\epsilon$ are implicit in model (\ref{eq:model}) and omitted there for 
    brevity. Here they denote the Gaussian, centred errors. That is, we have
    $y = \bm{\beta}^{\text{T}} \bm{\lambda} + \epsilon$, 
    $\mathbf{x} = \mathbf{B}^{\text{T}} \bm{\lambda} + \bm{\delta}$,
    with $\epsilon \sim \mathcal{N}(0,\sigma^2)$ and
    $\bm{\delta} \sim \mathcal{N}_p(0,\bm{\Psi})$.}
    \label{fig:model}
  \end{figure}

	In some applications, the features naturally come partitioned into groups
	$\mathcal{G}_1, \dots, \mathcal{G}_G$. Examples of such partitions are
	distinct functional networks of genes, features with significant versus 
	features with non-significant association to the outcome in a previous study, 
	and feature groups based on prior expert knowledge of feature importance
	\cite[see, e.g.,][]{munch_adaptive_2019}.
	Figure \ref{fig:model} displays model (\ref{eq:model}) with partitioned 
	features as a Bayesian network.
  
	Straightforward inclusion of the partitioning is possible through
	(i) a factor loadings partitioning, i.e., 
	$\mathbf{B} = \begin{bmatrix}\begin{array}{c|c|c}\mathbf{B}_1 & \cdots & 
	\mathbf{B}_G\end{array}\end{bmatrix}$, or (ii) a uniquenessess partitioning,
	i.e., 
	$$
	\bm{\Psi} = \begin{bmatrix}
	  \bm{\Psi}_1 &  & \\
	  & \ddots & \\
	  & & \bm{\Psi}_G
	\end{bmatrix}.
	$$
	Prediction (\ref{eq:prediction}) shows that the effect of
	inflation of diagonal element $\psi_{j}$ or shrinkage of column 
	$\mathbf{b}_{j}$ on the induced regression coefficients 
	$\tilde{\bm{\beta}}$ and $\bm{\beta}^*$, is similar. 
	With an appropriate choice of priors for $\mathbf{B}$ and 
	$\psi_j, \dots, \psi_p$, the partitioning of the 
	features $\mathcal{G}_1, \dots, \mathcal{G}_G$ is included in the prior model.
	Here, We pursue option (i) and model the feature structure through the 
	$\mathbf{B}$ prior, by considering groupwise constant 
	(up to a uniquenessess scaling) prior variances, i.e.,
	$\forall j \in \mathcal{G}_g: \gamma_j=\gamma_g$. The groupwise constant prior
	variance shrinks feature effects in the same group similarly. A small
	group-specific prior variance results in more shrinkage of feature effects,
	compared to a group with larger group-specific prior variance. By setting the
	group variances, the prior expected relevance of the group's features is
	encoded in the model. However, determining the prior variances is not 
	straighforward in most applications. Section \ref{sec:hyperparameters} 
	proposes an empirical Bayes approach to estimate these variances.
	
	\section{Estimation}\label{sec:estimation}
	Maximum likelihood estimation of model (\ref{eq:model2}) is straightforward
	when $n < \bar{p}$ and many algorithms are available in literature. 
	In the $\bar{p} > n$ domain, estimation is possible through penalized 
	likelihood maximisation. In the current paper, focus is on the Bayesian model,
	so we refer the reader to Sections \ref{sec:maximumlikelihood} and
	\ref{sec:penalizedmaximumlikelihood} of the SM for details on 
	maximum (penalized) likelihood estimation of (\ref{eq:model2}). 
	
	Bayesian posteriors are commonly approximated through MCMC sampling. 
	Sampling from the posterior of model 
	(\ref{eq:model2}) and (\ref{eq:prior}) is relatively straightforward 
	(see SM Section \ref{sec:gibbssampling} for a Gibbs sampler).
	However, due to high-dimensionality of the parameters, sampling is relatively 
	slow. In addition, the MCMC chain shows poor mixing in all investigated 
	applications and simulations. Poorly mixing MCMC chains require 
	a prohibitive number of samples to properly explore the posterior. Here, we
	avoid computationally expensive MCMC sampling by a mean-field variational 
	Bayes approximation of the posterior.
	
	Mean-field variational Bayes methods minimise the Kullback-Leibler divergence
	of the posterior from the (approximate) variational posterior. With observed 
	variables $\mathbf{X}$, some partitioning of 
  unobserved variables $\bm{\theta} = \{ \theta_1, \dots, \theta_K \}$, and
  an factorised posterior assumption $p(\bm{\theta} | \mathbf{X}) \approx
  \prod_{k=1}^K q(\theta_k)$, this results in marginal posteriors
  $q(\theta_k) \propto
  \exp \{ \mathbb{E}_{\bm{\theta}_{-k} | \theta_k, \mathbf{X}} 
  \[ \log p( \theta_k | \bm{\theta}_{-k}, \mathbf{X}) \] \}$. Note that we 
  slightly abuse notation and let $q(\cdot)$ denote distinct densities based 
  on the corresponding argument.
  For $p( \theta_k | \bm{\theta}_{-k}, \mathbf{X})$ in 
  the exponential family, $q(\theta_k)$ is in the same exponential
  family with natural parameter $\mathbb{E}_{\bm{\theta}_{-k} | \theta_k, 
  \mathbf{X}} \[ \eta (\bm{\theta}_{-k},\mathbf{X}) \]$, where 
  $\eta (\bm{\theta}_{-k},\mathbf{X})$ is the natural parameter of the full
  conditional distribution \cite[]{blei_variational_2017}.
  
  Here, let $\bm{\Lambda} = \begin{bmatrix} \bm{\lambda}_1 & \cdots 
  \bm{\lambda}_n \end{bmatrix}^{\text{T}}$ and the approximate posterior 
  factorise as
  \begin{equation}\label{eq:variationalposterior}
    p(\bm{\Lambda},\bar{\mathbf{B}},\bar{\psi}_1, \dots,
    \bar{\psi}_{\bar{p}} | \bar{\mathbf{X}}) \approx q(\bm{\Lambda})
    q(\bar{\mathbf{B}})q(\bar{\psi}_1, \dots, \bar{\psi}_{\bar{p}}),
  \end{equation}
  so that the approximate posterior that minimises the Kullback-Leibler 
  divergence of posterior to approximation is
  \begin{subequations}\label{eq:variationaldistributions}
    \begin{align}
      q(\bm{\Lambda}) & \overset{D}{=} \prod_{i=1}^{n} \mathcal{N}_d
      (\bm{\phi}_i, \bm{\Xi}), \\
      q(\bar{\mathbf{B}}) & \overset{D}{=} \prod_{\bar{j}=1}^{\bar{p}}
      \mathcal{N}_d (\bm{\mu}_{\bar{j}}, \bm{\Omega}_{\bar{j}}),\\
      q(\bar{\psi}_1, \dots, \bar{\psi}_{\bar{p}}) & \overset{D}{=}
      \prod_{\bar{j}=1}^{\bar{p}} \Gamma^{-1}
      (n/2 + d/2 + \kappa_{\bar{j}}, \zeta_{\bar{j}}).
    \end{align}
  \end{subequations}
  The so-called variational parameters are
  \begin{subequations}\label{eq:variationalparameters}
    \begin{align}
      \bm{\phi}_i & = \left\{ \sum_{{\bar{j}}=1}^{\bar{p}} \mathbb{E}
        (\bar{\psi}_{\bar{j}}^{-1})
        \[\mathbb{V}(\bar{\mathbf{b}}_{\bar{j}}) + 
        \mathbb{E}(\bar{\mathbf{b}}_{\bar{j}})
        \mathbb{E}(\bar{\mathbf{b}}^{\text{T}}_{\bar{j}})\] + 
        \mathbf{I}_d \right\}^{-1}
        \mathbb{E}(\bar{\mathbf{B}}) \mathbb{E}(\bar{\bm{\Psi}}^{-1})
        \bar{\mathbf{x}}_i, \, i=1, \dots, n,\\
      \bm{\Xi} & = \left\{ \sum_{{\bar{j}}=1}^{\bar{p}} 
        \mathbb{E}(\bar{\psi}_{\bar{j}}^{-1})
        \[\mathbb{V}(\bar{\mathbf{b}}_{\bar{j}}) + 
        \mathbb{E}(\bar{\mathbf{b}}_{\bar{j}})
        \mathbb{E}(\bar{\mathbf{b}}_{\bar{j}}^{\text{T}})\] + \mathbf{I}_d 
        \right\}^{-1},\\
      \bm{\mu}_{\bar{j}} &= \[\mathbb{E}(\bm{\Lambda}^{\text{T}})
        \mathbb{E}(\bm{\Lambda}) + n \mathbb{V}(\mathbf{\bm{\lambda}}_i) +
        \gamma_{{\bar{j}}}^{-1} \mathbf{I}_d\]^{-1}
        \mathbb{E}(\bm{\Lambda}^{\text{T}})\bar{\mathbf{x}}_{\bar{j}}, \, 
        {\bar{j}}=1,\dots, \bar{p},\\
      \bm{\Omega}_{\bar{j}} & = \mathbb{E}(\psi_{\bar{j}}^{-1})^{-1}
        \[\mathbb{E}(\bm{\Lambda}^{\text{T}})
        \mathbb{E}(\bm{\Lambda}) + n \mathbb{V}(\mathbf{\bm{\lambda}}_i) +
        \gamma_{{\bar{j}}}^{-1} \mathbf{I}_d\]^{-1}, \, {\bar{j}}=1,\dots 
        {\bar{p}}, \\
      \zeta_{\bar{j}} & = \bar{\mathbf{x}}_{\bar{j}}^{\text{T}} 
        \bar{\mathbf{x}}_{\bar{j}}/2 -
        \mathbb{E}(\bar{\mathbf{b}}_{\bar{j}}^{\text{T}})\mathbb{E}
        (\bm{\Lambda}^{\text{T}})
        \bar{\mathbf{x}}_{\bar{j}} + 
        \text{tr}\[\mathbb{E}(\bm{\Lambda}^{\text{T}})
        \mathbb{E}(\bm{\Lambda})\mathbb{V}(\bar{\mathbf{b}}_{\bar{j}})\]/2 +
        n\text{tr}\[\mathbb{V}(\bm{\lambda}_i)
        \mathbb{V}(\bar{\mathbf{b}}_{\bar{j}})\]/2 \nonumber \\
      & \,\,\,\,\,\, \,\,\, + \mathbb{E}(\bar{\mathbf{b}}_{\bar{j}}^{\text{T}})
        \mathbb{E}(\bm{\Lambda}^{\text{T}}) \mathbb{E}(\bm{\Lambda})
        \mathbb{E}(\bar{\mathbf{b}}_{\bar{j}})/2 + 
        n\mathbb{E}(\bar{\mathbf{b}}_{\bar{j}}^{\text{T}}) 
        \mathbb{V}(\bm{\lambda}_i)
        \mathbb{E}(\bar{\mathbf{b}}_{\bar{j}})/2 + 
        \gamma_{\bar{j}}^{-1}\mathbb{E}
        (\mathbf{b}_{\bar{j}}^{\text{T}})\mathbb{E}(\mathbf{b}_{\bar{j}})/2 
        \label{eq:zeta} \\
      & \,\,\,\,\,\, \,\,\, + \gamma_{\bar{j}}^{-1}
        \text{tr}\[\mathbb{V}(\mathbf{b}_{\bar{j}})\]/2 + \nu_{\bar{j}}, 
        \, {\bar{j}}=1,\dots,{\bar{p}}, \nonumber
    \end{align}
  \end{subequations}
  where we slightly abuse notation and let 
  $\bar{\mathbf{x}}_i$ and
  $\bar{\mathbf{x}}_{\bar{j}}$ denote the $i$th row and $\bar{j}$th column of 
  $\bar{\mathbf{X}}$, respectively.
  The expectations and variances are
  \begin{align*}
    \mathbb{E}(\bar{\psi}_{\bar{j}}^{-1}) & = (n/2 + d/2 + \kappa_{\bar{j}})/
      \zeta_{\bar{j}}, \, {\bar{j}}=1,\dots,{\bar{p}},\\
    \mathbb{E}(\bar{\mathbf{b}}_{\bar{j}}) & = \bm{\mu}_{\bar{j}}, 
      \, {\bar{j}}=1,\dots,{\bar{p}},\\
    \mathbb{V}(\bar{\mathbf{b}}_{\bar{j}}) & = \bm{\Omega}_{\bar{j}},
      \, {\bar{j}}=1,\dots,{\bar{p}},\\
    \mathbb{E}(\bm{\Lambda}) & = \bm{\Phi}, \\
    \mathbb{V}(\bm{\lambda}_i) & = \bm{\Xi}, \, i=1,\dots,n,
  \end{align*}
  with $\bm{\Phi} = \begin{bmatrix} \bm{\phi}_1 & \cdots \bm{\phi}_n
  \end{bmatrix}^{\text{T}}$. 
  The parameters contain cyclic dependencies and are
  updated until convergence.
  
	Model (\ref{eq:model}) describes a general covariance matrix. However, 
	a correlation matrix better describes the standardised data. In 
	the frequentist setting the general covariance model is straightforward to 
	extend to the correlation model by restriction of the likelihood to the space 
	of correlation matrices. Moreover, this
	is the default setting in the \texttt{R} package \texttt{factanal}.
	In the Bayesian setting, this requires either more intricate prior modelling 
	or post hoc corrections of the posterior. Here, we consider a post hoc
	correction that ensures that the posterior expectation of the covariance of
	$\bar{\mathbf{X}}$ desribes a correlation matrix: 
	$$
	\forall \bar{j}: 
	\E_{\bar{\mathbf{B}},\bar{\psi}_1 ,\dots, \bar{\psi}_{\bar{p}} | 
  \bar{\mathbf{X}}} \(\bar{\mathbf{b}}_{\bar{j}}^{\text{T}} 
  \bar{\mathbf{b}}_{\bar{j}} + {\bar{\psi}}_{\bar{j}}\) = 1.
  $$
	SM Section 
	\ref{sec:correlation} contains more details on 
	this posterior correction and a possible future direct correlation modelling
	approach. In the following, the post hoc correction approach is 
	applied.
	
	\subsection{Unlabeled observations}
	Inspection of (\ref{eq:prediction}) learns that the predictions 
	$\mathbb{E}(\tilde{y}|\tilde{\mathbf{x}})$
	depend on the observational model for $\mathbf{x}$ through $\mathbf{B}$ and
	$\bm{\Psi}$. As detailed in \cite{liang_use_2007}, this implies that 
	estimation benefits from additional unlabeled features $\mathbf{x}_i$,
	$i=n+1, \dots, n + m$, with the corresponding unobserved outcomes
	$z_i$, $i=n+1, \dots n + m$. A straightforward method of including the 
	unlabeled observations is to consider the full data likelihood
	$p(\mathbf{X}, \mathbf{z}, \mathbf{y} | \bar{\mathbf{B}}, \bar{\bm{\Psi}})$,
	with $\mathbf{z} = \begin{bmatrix} z_{n+1} & \cdots & z_{n+m} 
	\end{bmatrix}^{\text{T}}$
	\cite[]{banbura_maximum_2014,liu_maximum_1998}.
	Maximum likelihood estimation then requires marginalisation over unobserved
	outcomes $z_i$. Section \ref{sec:unlabeled} in the
	SM describes an EM algorithm for (penalized) maximum likelihood estimation 
	with missing data. 
	
	In the Bayesian model, the unobserved outcomes are now included in the 
	posterior distributions. The variational Bayes posterior 
	(\ref{eq:variationalposterior}) is augmented as
	$$
  p(\bm{\Lambda},\bar{\mathbf{B}},\bar{\psi}_1, \dots,
  \bar{\psi}_{\bar{p}}, \mathbf{z}| \bar{\mathbf{X}}) \approx q(\bm{\Lambda})
  q(\bar{\mathbf{B}})q(\bar{\psi}_1, \dots, \bar{\psi}_{\bar{p}})
  q(\mathbf{z}),
  $$
	where
	\begin{align*}
	  q(\mathbf{z}) & \overset{D}{=} \prod_{i={n+1}}^{n+m} \mathcal{N}
	    (\upsilon_i, \chi) \text{, with}\\
	  \upsilon_i & = \mathbb{E}(\bar{\mathbf{b}}_{\bar{p}}^{\text{T}} )
	    \mathbb{E}(\bm{\lambda}_i), \\
	  \chi & = \mathbb{E}(\bar{\psi}_{\bar{p}}^{-1})^{-1}.
	\end{align*}
	In addition, the term $\mathbbm{1}_{\bar{j}=\bar{p}}m\mathbb{V}(z_i)/2$ is 
	added to (\ref{eq:zeta}) and all occurences of $\bar{\mathbf{x}}_i$ and 
	$\bar{\mathbf{x}}_{\bar{j}}$ in (\ref{eq:variationalparameters}) are replaced
	with $\tilde{\mathbf{x}}_i$ and $\tilde{\mathbf{x}}_{\bar{j}}$, where
	$$
	\tilde{\mathbf{X}} = \begin{bmatrix}
    \multicolumn{1}{c}{\multirow{2}{*}{$\mathbf{X}$}} & \mathbf{y} \\
    \multicolumn{1}{c}{} & \mathbb{E}(\mathbf{z})
  \end{bmatrix},
	$$
	and
	\begin{align*}
	  \mathbb{E}(z_i) & = \upsilon_i, \\
	  \mathbb{V}(z_i) & = \chi.
	\end{align*}
	SM Section \ref{sec:gibbssampling} 
  contains more details on the inclusion of
  unlabeled observations in the (approximate) Bayesian posterior computations
  through MCMC. Although not shown here due to brevity, the 
  unobserved outcome approach is straightforward to extend to an unobserved
  features approach.  
	
	\subsection{Latent dimension}
	Although we initially assumed $d$ to be the true latent dimension, in general
	it is unkown and needs to be estimated. Methods for dimension estimation are 
	plentiful in the literature \cite[see, e.g., ][]{zwick_comparison_1986}. 
	Our modest aim of accurate prediction does not require correct estimation of 
	the latent dimension, as even picking the true latent dimension does not 
	always lead to optimal predictions \cite[]{goeman_statistical_2006}.
	Without this requirement of correct latent dimension estimation, we resort to
	the simple and fast Kaiser criterion. The Kaiser criterion picks $d$ 
	that retains
	dimensions with variance contribution larger than that of the average feature 
	$\mathbf{x}$. This amounts to choosing $d=\sum_{j=1}^p\mathbbm{1}
	\{v_j > 1\}$, with $v_j$, $j=1, \dots, p$, the eigenvalues of the correlation 
	matrix. That is, we set $d$ to the number of eigenvalues of the 
	correlation matrix of $\mathbf{X}$ larger than one.
	
	% In fact, 
	% \cite{goeman_statistical_2006} show the balance between bias and variance
	% that leads to optimal prediction error is for $d$ 
	% somewhere between 0 and the true latent dimension.
	% Here, we consider a few estimation 
	% methods:
	% \begin{enumerate}
	%   \item Pick $d$ such that a 
	%     pre-specified proportion of the variance of
	%     $\mathbf{X}$ is explained with the latent factors from a simple
	%     factor analysis. \cite{ferrari_bayesian_2020} propose the proportion of
	%     explained variance 0.9. In practice this amounts to choosing $d$, such 
	%     that $\sum_{j=1}^d v_j/p > 0.9$, where $v_j$ is the $j$th largest eigen
	%     value of the correlation matrix of $\mathbf{X}$.
	%   \item The Kaiser criterion picks $d$ such that we retain dimensions with
	%     variance contribution larger than that of the average feature 
	%     $\mathbf{x}$. This amounts to choosing $d=\sum_{j=1}^p\mathbbm{1}
	%     \{v_j > 1\}$, i.e., set $d$ to the number of eigen values of the 
	%     correlation matrix of $\mathbf{X}$ larger than one.
	%   \item The Marchenku-Pastur law based rule picks $d$ such that all 
	%     dimensions contribute more to the variance of $\mathbf{X}$ than
	%     than of the average feature in the asymptotic limit 
	%     ($n,p \rightarrow \infty$) of a random matrix
	%     $\mathbf{X}_{n \times p}$. In practice, this means choosing 
	%     $d=\sum_{j=1}^p \mathbbm{1} \{ v_j > (1 + \sqrt{p/n})^2 \}$.
	% \end{enumerate}
	% 
	% - more explanation needed
	
	\subsection{Hyperparameters}\label{sec:hyperparameters}
	The Bayesian model requires a choice of
	hyperparameters $\gamma_g$, $\kappa_j$ and $\nu_j$. 
	Choosing the $\gamma_g$ by hand requires intricate prior expert
	knowledge, which might not be available. An alternative is to estimate them
	from the data using empirical Bayes. Or, if we do know the overall scale of
	the $\gamma_g$, but not the group-specific deviations, we may reparametrise as
	$\gamma_g = \gamma \gamma_g'$, fix the overall scale $\gamma$ and estimate the
	group-specific multipliers $\gamma_g'$.
	
	In both empirical Bayes settings we maximise the marginal 
	likelihood (constrained maximisation for the second approach). Direct 
	marginal likelihood maximisation requires calculation of a $p$-dimensional 
	integral for which no closed form is available. With $p$ large (i.e., the high
	dimensional setup considered here), an EM algorithm with iterations
	$$
  \bm{\gamma}^{(k + 1)} = \underset{\bm{\gamma}}{\argmax} 
  \mathbb{E}_{ \theta | \mathbf{y}} \[\log p(\bar{\mathbf{B}} | 
  \bar{\psi}_1, \dots, \bar{\psi}_{\bar{p}} ) | 
  \bm{\gamma}^{(k)} \],
  $$
	where $\bm{\gamma} = \begin{bmatrix} \gamma_1 & \cdots \gamma_G 
	\end{bmatrix}^{\text{T}}$, is computationally much more feasible.
	With a variational Bayes approximation of the difficult expectation, this
  results in
  $$
  \bm{\gamma}^{(k + 1)} = \underset{\bm{\gamma}}{\argmax}\left\{
  - \frac{1}{2} \sum_{g=1}^G \gamma_g^{-1} 
  \sum_{j \in \mathcal{G}_g} \mathbb{E}(\bar{\psi}_j^{-1}) 
  \left\{ \text{tr} \[ 
  \mathbb{V}(\bar{\mathbf{b}}_j)\] + \mathbb{E}(\bar{\mathbf{b}}_j^{\text{T}})
  \mathbb{E}(\bar{\mathbf{b}}_j) \right\} - \frac{d}{2} 
  \sum_{g=1}^{G} |\mathcal{G}_g| \log \gamma_{g} \right\}.
  $$
  Finally, this gives empirical Bayes updates:
  $$
  \gamma_g^{(k + 1)} = \frac{\sum_{j \in \mathcal{G}_g} 
  \mathbb{E}(\bar{\psi}_j^{-1}) \left\{ \text{tr} \[ 
  \mathbb{V}(\bar{\mathbf{b}}_j)\] + \mathbb{E}(\bar{\mathbf{b}}_j^{\text{T}})
  \mathbb{E}(\bar{\mathbf{b}}_j) \right\}}{|\mathcal{G}_g| d}.
  $$
  For the $\gamma_g = \gamma \gamma_g'$ parametrisation, the updates
  \begin{align*}
  \bm{\gamma}'^{(k + 1)} & = \underset{\bm{\gamma}'}{\argmax}\left\{ 
  - \frac{1}{\gamma} \sum_{g=1}^G \gamma_g'^{-1} 
  \sum_{j \in \mathcal{G}_g} \mathbb{E}(\bar{\psi}_j^{-1}) 
  \left\{ \text{tr} \[ 
  \mathbb{V}(\bar{\mathbf{b}}_j)\] + \mathbb{E}(\bar{\mathbf{b}}_j^{\text{T}})
  \mathbb{E}(\bar{\mathbf{b}}_j) \right\} - \frac{d}{2} 
  \sum_{g=1}^{G} |\mathcal{G}_g| \log \gamma_{g}' \right\}, \\
  & \text{subject to } \prod_{g=1}^G \gamma_g'^{|\mathcal{G}_g|}=1,
  \end{align*}
  are not available
  in closed form, but still convex and easy to compute with standard numerical
  optimisation tools.
	Empirical Bayes estimation of the $\gamma_g$ or $\gamma_g'$ is data-dependent 
	and does not rely on subjective arguments. In addition, empirical Bayes 
	estimation avoids (possibly complicated) hyperpriors on the $\gamma_g$
	and $\gamma_g'$. A 
	drawback is that we lose the uncertainty propagation property of the full 
	Bayesian approach.
	
	Prior error variance/uniquenesses shapes $\kappa_{\bar{j}}$ and scale 
	$\nu_{\bar{j}}$, and overall prior variance $\gamma$ are set to default values
	to reflect a lack of prior 
	knowledge. Our default choice of hyperparameters should take the
	standardisation
	of the data into account. Three postulates are used to select the
	hyperparameters:
	(i) we ensure that the prior expectation describes
	a correlation matrix model, i.e., 
	$\forall j: \E_{\bar{\mathbf{B}},\bar{\psi}_1
	,\dots, \bar{\psi}_{\bar{p}}} (\bar{\mathbf{b}}_{\bar{j}}^{\text{T}} 
  \bar{\mathbf{b}}_{\bar{j}} + {\bar{\psi}}_{\bar{j}}) = 1$. 
  Furthermore, (ii) the prior contributions of the error and the latent 
  structure to the data are assumed equal, i.e., 
  $\forall j: \E_{\bar{\mathbf{B}},\bar{\psi}_1
	,\dots, \bar{\psi}_{\bar{p}}} (\bar{\mathbf{b}}_{\bar{j}}^{\text{T}} 
  \bar{\mathbf{b}}_{\bar{j}}) = \E_{\bar{\mathbf{B}},\bar{\psi}_1
	,\dots, \bar{\psi}_{\bar{p}}} ({\bar{\psi}}_{\bar{j}}) = 1/2$.
  Lastly, (iii) the prior uniqueness variance is set to 
  $\mathbb{V}_{\bar{\mathbf{B}},\bar{\psi}_1
	,\dots, \bar{\psi}_{\bar{p}}} ({\bar{\psi}}_{\bar{j}}) = 1$.
	These three postulates together result in $\gamma = 1/d$, 
	$\forall \bar{j}: \kappa_{\bar{j}}=9$, and 
	$\forall \bar{j}: \nu_{\bar{j}}=4$. As a result,
	$\forall {\bar{j}}: \mathbb{V}_{\bar{\mathbf{B}},\bar{\psi}_1
	,\dots, \bar{\psi}_{\bar{p}}}(\mathbf{b}_j^{\text{T}}\mathbf{b}_j) = 1 + 
	5/(2d)$. For $d$ large compared to $5/2$ (as one expects in high-dimensional
	settings), we have $\mathbb{V}_{\bar{\mathbf{B}},\bar{\psi}_1
	,\dots, \bar{\psi}_{\bar{p}}}(\mathbf{b}_j^{\text{T}}\mathbf{b}_j) \approx
	1 = \mathbb{V}_{\bar{\mathbf{B}},\bar{\psi}_1
	,\dots, \bar{\psi}_{\bar{p}}} ({\bar{\psi}}_{\bar{j}})$, so that the 
	contributions to 
	the prior variance of latent structure and error are approximately equal.
	The methods described in this Section are implemented in the \texttt{R}
	package \texttt{bayesfactanal} available from 
	\url{https://github.com/magnusmunch/bayesfactanal}.
	
	\section{Simulations}\label{sec:simulations}
	To assess the potential benefit of the proposed models in prediction of
	outcome $y$ from features $\mathbf{x}$, 
	a simulation is set up. The simulation setting is 
	meant to demonstrate the potential benefit of (i) the Bayesian factor 
	regression model in general, (ii) the inclusion of the feature structure 
	through the empirical Bayes estimation of the $\gamma_g'$ as explained in 
	Section \ref{sec:hyperparameters}, and (iii) the use of unlabeled features in
	the estimation.
	
	To that end, $n=50$ labeled, 
	and $m \in \{0, 50, 100, 200, 500 \}$ unlabeled observations are drawn from 
	model (\ref{eq:model}) and standardised after simulation. 
	Error variance and uniquenesses are set to 
	$\sigma^2=1$ and $\forall j: \psi_j=1$. The number of features is fixed to
	$p=100$. Two scenarios for the model parameters $b_{hj}$ and $\beta_h$ are 
	considered:
	\begin{enumerate}
	  \item the number of factors are fixed to $d=10$. The $b_{hj}$ are set so 
	    that each feature loads on two factors and factor is a part of 20
	    features (see (\ref{eq:simulation}), where each $b$ denotes
	    10 values and the empty cells are set to zero). $\beta_h$ is set so that 
	    the outcome loads on all factors. 
	    The features are divided in two groups 
	    $\mathcal{G}_1 = \{1, \dots, 50 \}$ and 
	    $\mathcal{G}_2 = \{51, \dots, 100 \}$.
	    The non-zero $b_{hj}$ are 
	    drawn from independent univariate centered Gaussian distributions. The 
	    variances are $\mathbb{V}(b_{hj})=0.1$ for $j=1, \dots 50$, and
	    $\mathbb{V}(b_{hj})=1$ for $j=51, \dots, 100$ The $\beta_h$ are 
	    independent and drawn from the standard Gaussian distribution. 
	  \item The second scenario fixes $d=40$. The model parameters $b_{hj}$ are 
	    drawn from independent univariate centered Gaussian distributions. The 
	    variances are $\mathbb{V}(b_{hj})=0.1$ for $j=1, \dots 50$, and
	    $\mathbb{V}(b_{hj})=10$ for $j=51, \dots, 100$, i.e., the features are
	    structured in two groups: $\mathcal{G}_1 = \{1, \dots, 50 \}$ and
	    $\mathcal{G}_2 = \{51, \dots, 100 \}$. To ensure that
	    the proportion of variance in $y$ explained with the factors is
	    0.7, $\beta_h$ is set $\beta_h=0.483$.
	\end{enumerate}
	\begin{equation}\label{eq:simulation}
	  \mathbf{B}=
	  \left[
      \begin{array}{ c c c c c:c c c c c}
        b & &  &  &  &  &  &  &  & b\\ 
        b & b &  &  &  &  &  &  &  & \\
        & b & b &  &  &  &  &  &  & \\
        &  & b & b & &  &  &  &  & \\
        &  &  & b & b & &  &  &  & \\
        &  &  &  & b & b &  &  &  & \\
        &  &  &  &  & b & b &  &  & \\
        &  &  &  &  &  & b & b & & \\
        &  &  &  &  &  &  & b & b & \\
        &  &  &  &  &  &  &  & b & b 
      \end{array} 
    \right]
  \end{equation}
	The first scenario models a situation where the features load on two factors 
	only in such a way that the marginal correlation between features is weak. 
	This might occur, for example, if genes are 
	organised in
	nearly disjoint functional networks, but the outcome is related to all the
	networks. Ridge regression is expected to 
	perform well here. With such a sparse loadings matrix, 
	$(\mathbf{B}^{\text{T}}\mathbf{B} + \bm{\Psi})^{-1} \approx \mathbf{I}_p$.
	That is, the information in $\mathbf{X}$ contributes little to the induced
	regression coefficients $\tilde{\bm{\beta}}$. In addition, the induced
	regression coefficients become
	$\tilde{\bm{\beta}} \approx \mathbf{B}^{\text{T}} \bm{\beta} = 
	\mathbb{C}\text{ov}(y, \mathbf{x})$, a (rescaled version of the) quantity 
	that standard linear regression methods aim to estimate.
	
	The second scenario models a setting where all features load on all factors,
	but the strength of the loading depends on the feature group. This might 
	occur, for example, if genes are organised in several interconnected 
	functional networks, but some network have weak connections. The outcome is 
	again related to the all functional networks. In this setting, the
	factor regression methods are expected to perform well. In contrast to the
	first simulation, 
	$(\mathbf{B}^{\text{T}}\mathbf{B} + \bm{\Psi})^{-1} \neq \mathbf{I}_p$,
	so information on the induced regression coefficients $\tilde{\bm{\beta}}$ is
	contained in $\mathbf{X}$. This results in increased effiency due to the 
	inclusion of data. Also, $\tilde{\bm{\beta}}$, is a weighted version of
	$\mathbb{C}\text{ov}(y, \mathbf{x})$ that is not straightforward to estimate
	with standard linear regression methods.
	
	Six models are compared:
	\begin{enumerate}
	  \item Ridge regression with cross validated penalty parameter with the
	    \texttt{R} package \texttt{glmnet} \cite[]{friedman_regularization_2010};
	  \item Lasso regression with cross validated penalty parameter with the
	    \texttt{R} \texttt{glmnet} package \cite[]{friedman_regularization_2010};
	  \item a two-step factor regression method: (i) a penalized factor 
	    model is estimated from the feature correlation matrix, with cross 
	    validated penalty parameter.
	    Next, (ii) outcomes are regressed on the feature factor scores 
	    $\hat{\mathbb{E}}(\bm{\lambda}_i | \mathbf{x}_i)$ to obtain the prediction 
	    rule. This approach was shown to work in \cite{peeters_stable_2019} and
	    is implemented in the \texttt{R} \texttt{FMradio} package 
	    \cite[]{peeters_fmradio_2019};
	  \item a penalized factor regression model that includes unlabeled 
	    observations, with cross validated penalty parameter, and estimated as in
	    SM Section (\ref{sec:mlestimation});
	  \item the proposed Bayesian factor regression model (\ref{eq:prior}), 
	    approximated with variational Bayes as in Section \ref{sec:estimation}. 
	    The fixed hyperparameters are described in Section 
	    \ref{sec:hyperparameters}. Note that this model does not include 
	    external feature structure and therefore does not estimate the 
	    $\gamma_g'$;
	  \item the proposed empirical Bayesian factor regression model 
	    (\ref{eq:prior}), approximated with variational Bayes as in Section 
	    \ref{sec:estimation}. The hyperparameters are described in Section 
	    \ref{sec:hyperparameters}, where we include the grouping of the features
	    and estimate group-specific $\gamma_g'$ by empirical Bayes.
	\end{enumerate}
	For all models, the data are standardised before estimation, as is common
	in most real data applications.
	Models 3-6 allow for the inclusion of unlabeled features and are estimated
	for a range of number of unlabeled features.
	In addition, we 
	% consider the true data-generating model $\tilde{\beta}$ and
	fitted an intercept-only null model. We calculate estimation mean squared 
	error
	(EMSE) of $\tilde{\bm{\beta}}$, prediction mean squared error (PMSE),
	and correlation between predictions and observations 
	($\mathbb{C}\text{or}(y,\hat{y})$) on test data of size 
	$n_{\text{test}}=1000$. Lower PMSE and EMSE indicate better performance, while
	higher $\mathbb{C}\text{or}(y,\hat{y})$ indicates better performance.
	The results, with the median taken
	over 50 simulation replications, are displayed in Figures
	\ref{fig:simulation1} and Figures 
	\ref{fig:simulation2}, for scenarios 1 and 2, respectively.
	
\begin{knitrout}
\definecolor{shadecolor}{rgb}{0.969, 0.969, 0.969}\color{fgcolor}\begin{figure}[!ht]

{\centering \includegraphics[width=1\linewidth]{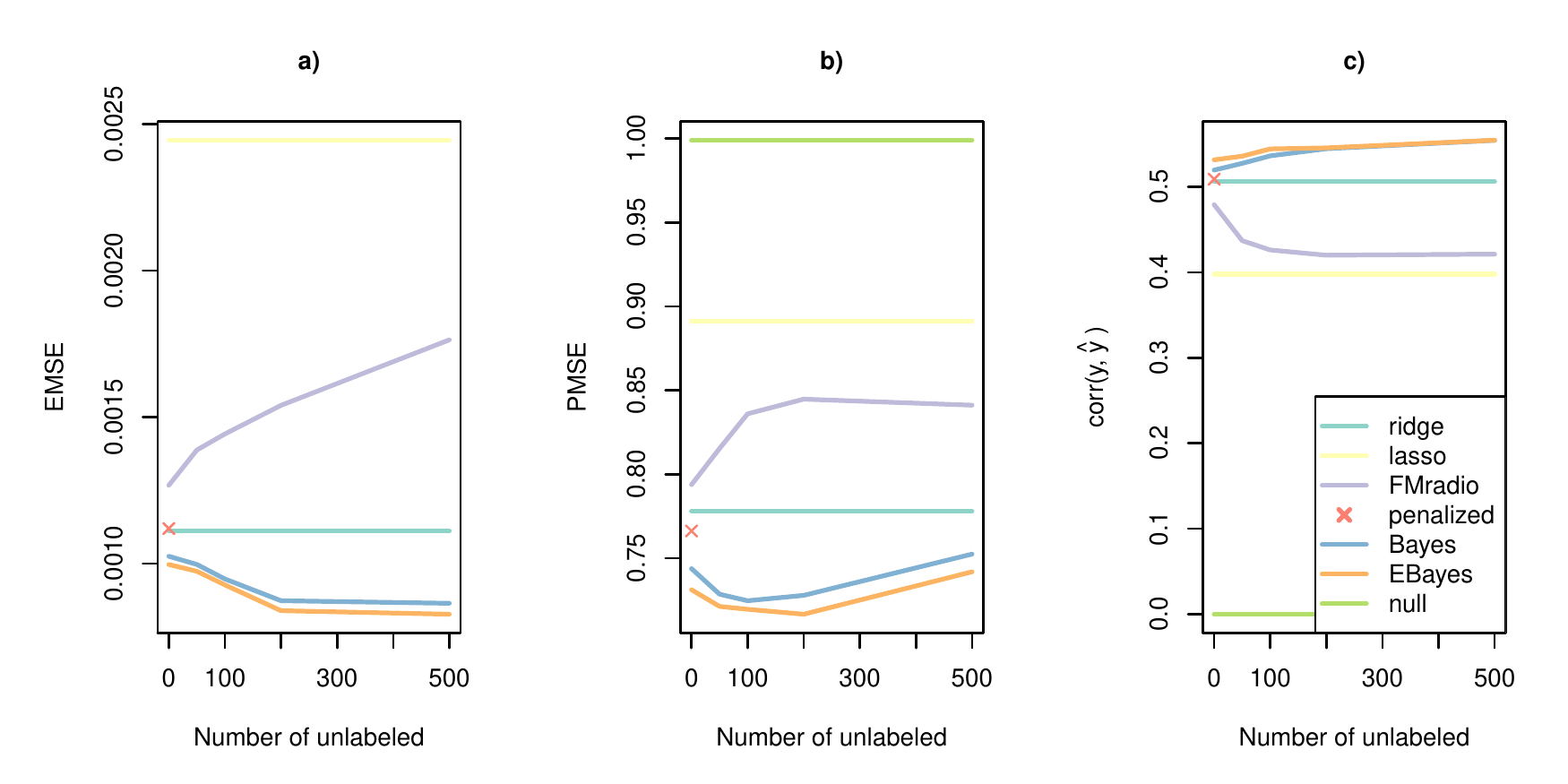} 

}

\caption[Simulation results for scenario 1 with median (a) EMSE and (b) PMSE, respectively]{Simulation results for scenario 1 with median (a) EMSE and (b) PMSE, respectively.}\label{fig:simulation1}
\end{figure}

\end{knitrout}
\begin{knitrout}
\definecolor{shadecolor}{rgb}{0.969, 0.969, 0.969}\color{fgcolor}\begin{figure}[!ht]

{\centering \includegraphics[width=1\linewidth]{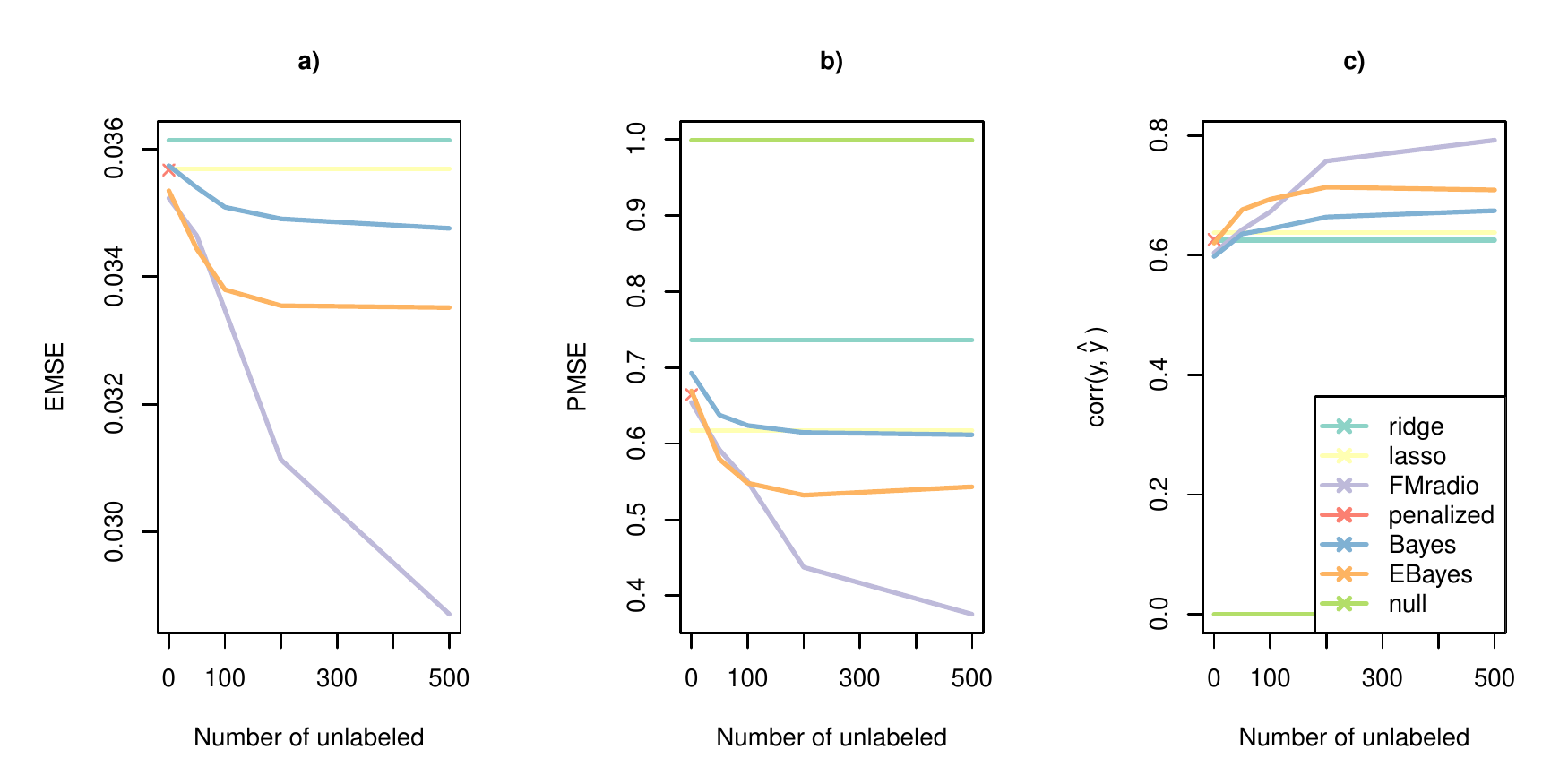} 

}

\caption[Simulation results for scenario 2 with median (a) EMSE and (b) PMSE, respectively]{Simulation results for scenario 2 with median (a) EMSE and (b) PMSE, respectively.}\label{fig:simulation2}
\end{figure}

\end{knitrout}

  In both scenarios, the penalized factor regression model was not estimable
  with unlabeled data, due to non-convergence. The two-step FMradio approach
  also suffers from non-convergence with more unlabeled data, but was still 
  estimable in some simulations, so it is included in the Figures.

  In both scenarios estimation (i.e., EMSE) and prediction calibration 
  (i.e., PMSE) of the Bayesian methods
  initially improves with more unlabeled data. However, in scenario 1 it starts
  deteriorating again after about $m=100$. Surprisingly, the opposite holds for 
  FMradio. In scenario 2, where the performance continues to improve with 
  more unlabeled data, the rate of improvement decreases with the number of
  unlabeled observations. This is unsurprising, as estimators generally converge
  at a similarly-shaped $\sqrt{n}$ rate. In both scenarios, discrimination 
  (i.e., $\mathbb{C}\text{or}(y,\hat{y})$) keeps improving with the addition
  of unlabeled features. For scenario 1, this is surprising, considering the
  eventual deterioration in calibration and estimation.
  
  In scenario 1, the Bayesian methods outperform the frequentist methods for
  almost all $m$ in terms of estimation and discrimination. Calibration is
  worse for the Bayesian methods for small and large $m$, but better for medium
  $m$. The two-step factor regression model FMradio, performs worse than the
  Bayesian factor regression methods and ridge, only outperforming lasso.
  In scenario 2, the frequentist methods outperform the Bayesian method for 
  small $m$ in terms of estimation and calibration.
  For medium $m$, the Bayesian methods outperform ridge, and eventally,
  for large $m$, also lasso. FMradio outperforms all other methods in 
  estimation, calibration, and discrimination. Scenario 2 simulates strong 
  factors, that explain much of the data. Extraction of these factors 
  in step one of the FMradio approach is therefore relatively easy. Estimation
  of the prediction rule based on these strong factors in step two of FMradio
  then results in a strong predictor.
  
  A comparison of full Bayes and empirical Bayes shows that the inclusion of 
  the feature groupings helps in both estimation and prediction. In scenario 1,
  empirical Bayes estimation and calibration is slightly better than full 
  Bayes. Discrimination is about equal. In scenario 2 empirical clearly
  outperforms full Bayes in all three performance measures.
  Figures
  \ref{fig:gamma1} and \ref{fig:gamma2} display the estimated 
  $\log \hat{\gamma}'_g$ for the empirical Bayes model in scenarios 1 and 2,
  respectively.
  Both Figures show a clear influence of the feature grouping
  on estimation, as the prior variances of the groups show a clear difference.
  Furthermore, the influence of the feature grouping grows with the number
  of unlabeled observations, as the diverging lines indicate.
\begin{knitrout}
\definecolor{shadecolor}{rgb}{0.969, 0.969, 0.969}\color{fgcolor}\begin{figure}[!ht]

{\centering \includegraphics[width=0.4\linewidth]{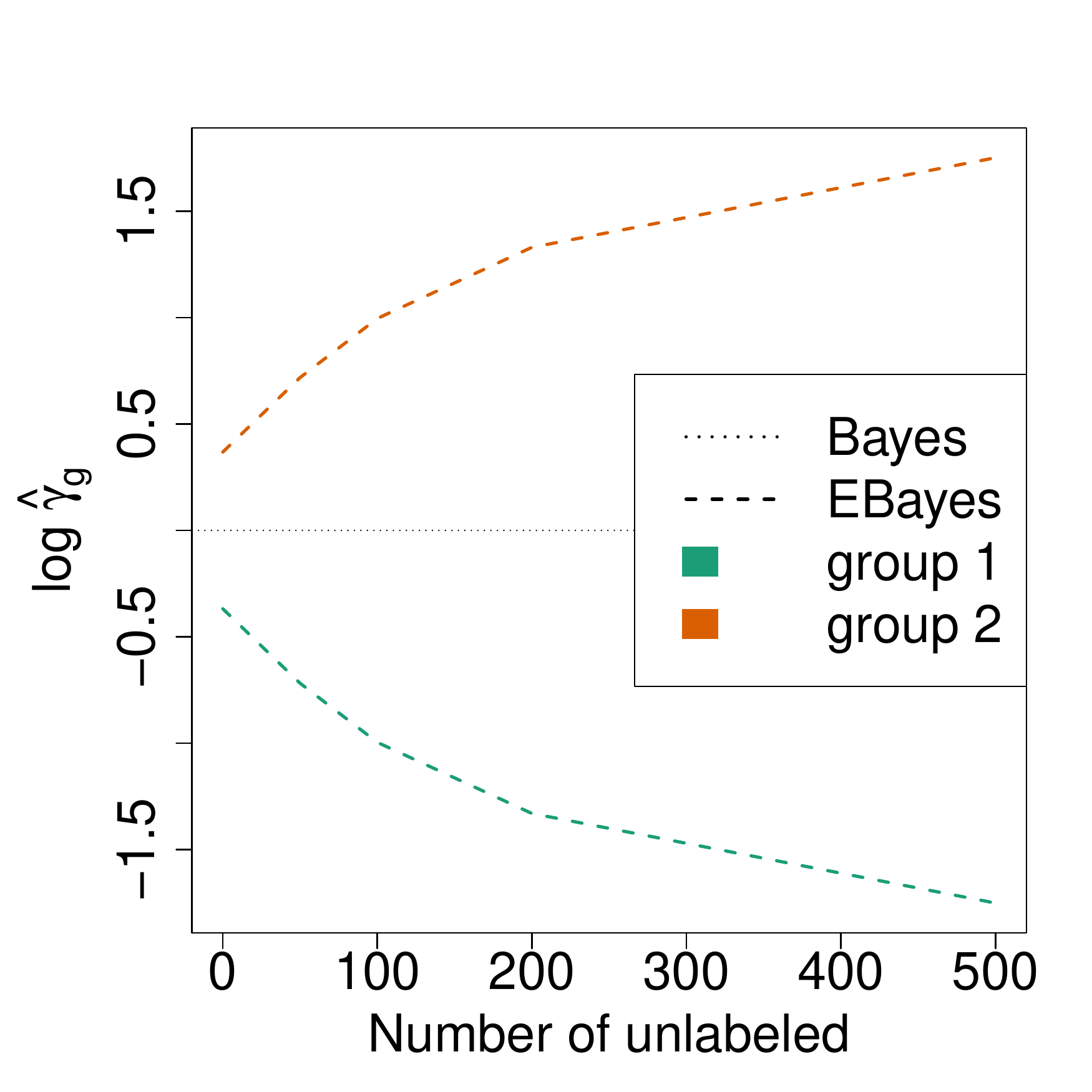} 

}

\caption[Simulation results for scenario 1 with median $\log \hat{\gamma}_g'$ estimated with empirical Bayes]{Simulation results for scenario 1 with median $\log \hat{\gamma}_g'$ estimated with empirical Bayes.}\label{fig:gamma1}
\end{figure}

\end{knitrout}
\begin{knitrout}
\definecolor{shadecolor}{rgb}{0.969, 0.969, 0.969}\color{fgcolor}\begin{figure}[!ht]

{\centering \includegraphics[width=0.4\linewidth]{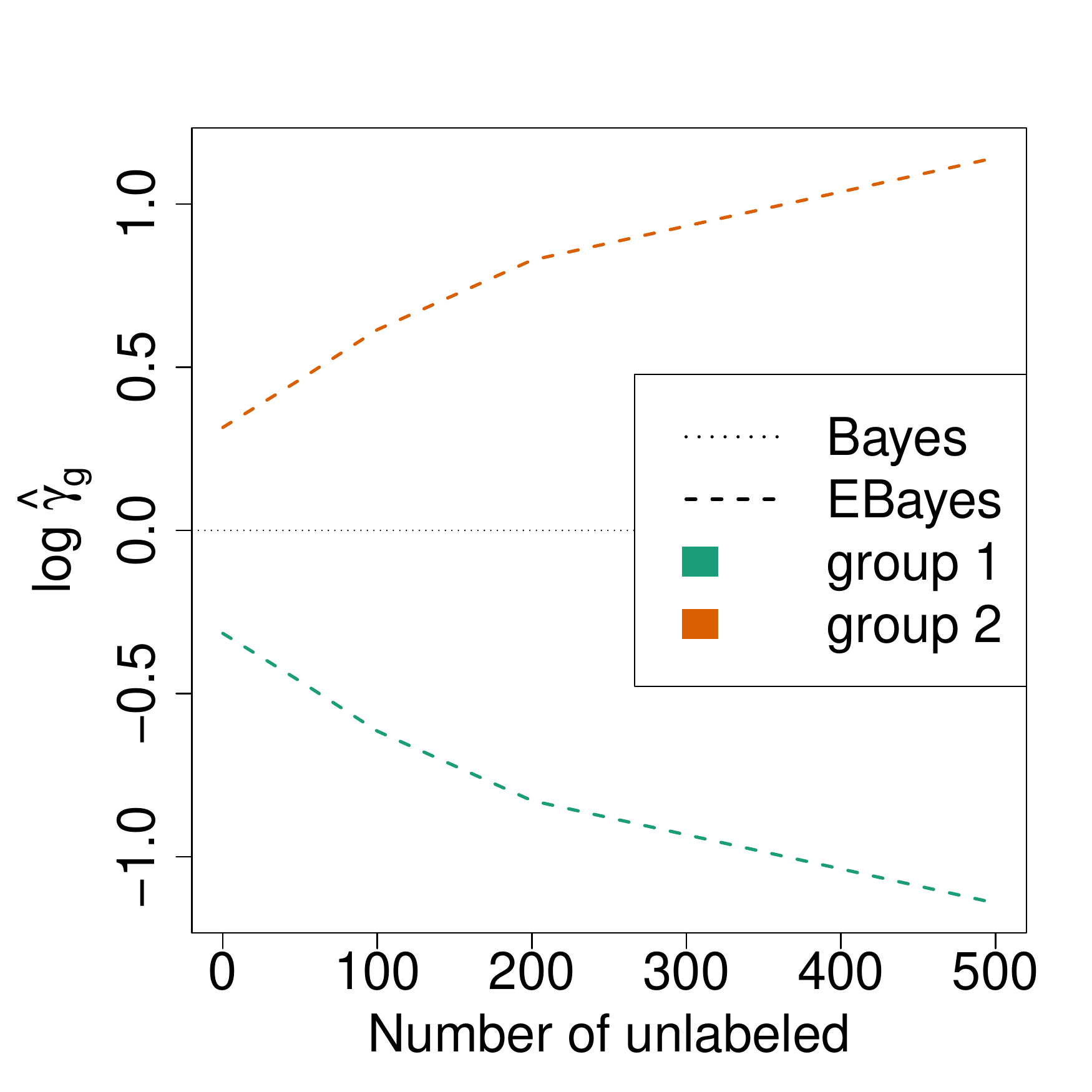} 

}

\caption[Simulation results for scenario 2 with median $\log \hat{\gamma}_g'$ estimated with empirical Bayes]{Simulation results for scenario 2 with median $\log \hat{\gamma}_g'$ estimated with empirical Bayes.}\label{fig:gamma2}
\end{figure}

\end{knitrout}

	\section{Applications}\label{sec:applications}
	\subsection{Influenza vaccine}
	The data described in this Section are from \cite{nakaya_systems_2011} and 
	made publicly available through the NCBI GEO archive 
	\cite[]{barrett_ncbi_2012} with accession numbers GSE29614 and GSE29617. 
	The analysis mostly follows \cite{van_deun_obtaining_2018}, 
	where a main aim was to predict vaccine efficacy with microarray gene
	expression data. Here follows a short description of the data; 
	for more details we refer the reader to
	\cite{van_deun_obtaining_2018}.
	
	The data are from 9 and 26 subjects, observed in the 2007 and 2008 flu 
	seasons, respectively. For all subjects there are three efficacy measures
	from just before and 28 after vaccination available in the form of three
	different plasma hemagglutination inhibition (HAI) antibody titers. The
	antibody titers were combined by first taking the maximum of the three 
	log-transformed titers and then substracting the measurements just before and
	three days after vaccination. The scores were
	standardised to mean zero and variance one.
	In addition to the vaccine efficacy measures, there are 54,675 microarray gene 
	expression measurements available from just before and three days after 
	vaccination. The Robust Multichip Average (RMA) algorithm 
	\cite[]{irizarry_exploration_2003} 
	was used to pre-process the microarrays. After pre-processing, a change 
	score was calculated by substracting the measurements just before and three
	days after vaccination from each other. These scores were standardised to
	mean zero and variance one. Before the analysis, a pre-selection
	of 416 genes with highest coefficient of variation is made. 
	The choice of 416 genes follows the analysis results of 
	\cite{van_deun_obtaining_2018}. Here, we consider the 2007 data as unlabeled
	and the 2008 data as labeled.  
	
	The application is an example of a difficult high-dimensional prediction 
	problem, with little data available: a situation that regularly arises in 
	practice. Here, the available unlabeled data potentially 
	increases predictive performance signficicantly. Additionally, genes are often
	considered to be organised in functional networks, so the factor model
	is an appropriate choice and we expect the factor regression methods to
	outperform classical linear regression methods.
	
	We estimate the same models as in Section \ref{sec:simulations}, 
	with the exception of the empirical Bayes model, because there is no
	grouping of the features available. To assess performance we
	calculated cross-validated PMSE and $\mathbb{C}\text{or}(y,\hat{y})$ 
	and display them in Table
	\ref{tab:application1}, where null refers to the intercept only model.
	The penalized factor regression model did not converge, so is not included 
	in the results.
\begin{knitrout}
\definecolor{shadecolor}{rgb}{0.969, 0.969, 0.969}\color{fgcolor}\begin{table}[H]

\caption{\label{tab:application1}Cross-validated PMSE and $\mathbb{C}\text{or}(y,\hat{y})$ (best performing in bold) calculated on the influenza vaccine data.}
\centering
\begin{tabular}[t]{lrr}
\toprule
  & PMSE & $\mathbb{C}\text{or}(y,\hat{y})$\\
\midrule
ridge & 0.959 & 0.171\\
lasso & 0.929 & 0.339\\
FMradio & 0.955 & 0.097\\
VBayes & \textbf{0.866} & \textbf{0.341}\\
null & 0.962 & 0\\
\bottomrule
\end{tabular}
\end{table}

\end{knitrout}
	Table \ref{tab:application1} shows that the variational Bayesian factor 
	regression that includes the unlabeled data outperforms the other methods in 
	terms of calibration (i.e., PMSE) and discrimination 
	(i.e., $\mathbb{C}\text{or}(y,\hat{y})$), according to expectation. The other
	methods perform similarly in terms of PMSE, while lasso performance approaches 
	the Bayesian factor regression in terms of $\mathbb{C}\text{or}(y,\hat{y})$.
	
	\subsection{Oral cancer lymph node metastasis}
	In this Section, oral cancer lymph node metastasis is predicted with gene
	expression data. RNAseqs, taken from TCGA 
	\cite[]{the_cancer_genome_atlas_network_comprehensive_2015}, are measured on 
	133 HPV-negative oral tumours taken from 76 and 57 oral cancer patients, with
	and without lymph node metastasis, respectively. 
	For more details on these data, see \cite{te_beest_improved_2017}.
	Additional gene expressions are available from an independent 
	microarray study on 97 oral cancer patients in
	\cite{mes_prognostic_2017}. 
	These microarrays are normalised to the same scale as
  the RNAseqs and included in the analysis as unlabeled data.
	A set of 871 genes with $p \leq 0.01$ in the microarray data is 
	pre-selected.
  To investigate the empirical Bayes estimation of the $\gamma_g$, the 
	genes are divided in three groups, based on the cis-correlation between
	between the RNAseq data and TCGA DNA copy numbers on the same patients, 
	quantified by Kendall's $\tau$. 
	% The procedure to group the features is
	% described in \cite{munch_adaptive_2019}. 
	
	This Section investigates an example of a high-dimensional classification
	problem in for which both unlabeled data and external feature
	information is available. As before, genes are assumed to be organised in
	functional networks, so we expect the factor regression methods to fit the
	data well. We expect features with a large
	positive correlation between RNAseqs and DNA copy number, as quantified by
	Kendall's $\tau$, to be more important for metastasis prediction. We
	therefore expect to estimate larger $\gamma_g'$ for the groups with
	higher Kendall's $\tau$.
	
	We estimate the logistic extensions of the models estimated in Section 
	\ref{sec:simulations}. To assess performance we
	calculated a calibration measure Brier skill score (BSS) and discrimination 
	measure area under the receiver operator curve (AUC)
	on the unlabeled data and display them in Table
	\ref{tab:application2}.
	The penalized factor regression model did not converge, so is not included 
	in the results.
\begin{knitrout}
\definecolor{shadecolor}{rgb}{0.969, 0.969, 0.969}\color{fgcolor}\begin{table}[H]

\caption{\label{tab:application2}BSS and AUC (best performing in bold) calculated on the oral cancer lymph node metastasis data.}
\centering
\begin{tabular}[t]{lrr}
\toprule
  & BSS & AUC\\
\midrule
ridge & 0.125 & 0.698\\
lasso & \textbf{0.132} & 0.708\\
FMradio & 0.014 & 0.66\\
VBayes & 0.099 & 0.746\\
EBayes & 0.101 & \textbf{0.748}\\
\bottomrule
\end{tabular}
\end{table}

\end{knitrout}
  Surprisingly, the best performing model in terms of calibration (i.e., BSS) 
  is the lasso. The Bayesian factor regression methods outperform the
  other methods in terms of discrimination (i.e., AUC).
  The estimated $\gamma_g'$ are $0.97$, $0.98$, and $1.01$ for the low-, 
  medium-, and high cis-correlation groups, respectively. This small difference
  in shrinkage leads to a marginal increase in predictive performance of the
  empirical Bayes method compared to the full Bayes version.
	% \subsection{Alzheimers disease}
	% - check Perrakis and Mukherjee (2019) data
	% - try metabolomics data
	
	\section{Discussion}\label{sec:discussion}
	This paper investigates a Bayesian factor regression model for high 
	dimensional prediction and classification problems. It allows for the 
	inclusion of unlabeled data and feature groupings to improve predictive
	performance. Estimation is through a combination of variational and
	empirical Bayes techniques. The approach is competitive with classical
	ridge and lasso regression, as well as with more elaborate frequentist
	factor modelling 
	approaches such as penalized factor regression and the two-step factor 
	\texttt{FMradio}. Simulations show that
	the method is especially useful if the features are generated in 
	dense, correlated networks. Two applications show that the method predicts
	just as well, or better, than existing methods in real data settings.
	
	A technical advantage of the pursued factor modelling approach is the 
	straightforward inclusion of unlabeled observations through the full 
	likelihood approach. However, some caution regarding this approach is 
	advised. For the full likelihood approach to return unbiased estimates,
	the missing data mechanism is assumed to be at most missing at random (MAR). 
	That is, the missingness possibly depends on the observed features, but not
	on unobserved features. In the current setting, MAR implies that unobserved 
	labels are not missing due to the value of the labels. We argue that in most 
	applications, this is a reasonable assumption. In the examples above,
	observations are unlabeled because they come from independent studies. 
	Due to the independence, it is reasonable to assume that no relation
	exists between not observing labels and the actual labels.
	Another technical advantage of Bayesian modelling is the occurence of
	convergence issues in frequentist models. Sections \ref{sec:simulations} and 
	\ref{sec:applications} 
	show that the frequentist factor models suffer from convergence 
  issues if the number of labeled and/or unlabeled samples becomes large. 
  More investigation is required to determine when and why these convergence
  issues occur.
  An inherent benefit of Bayesian modelling is the
  uncertainty quantification that automatically comes with the Bayesian 
  posterior. This allows for straightforward calculation of prediction 
  intervals. We note that the uncertainty quantification in the current setting
  requires a more thorough investigation.
	
	More elaborate prior modelling of the factor 
  loadings is possible through the $\gamma_j$. For example, a more sparse 
  lasso model for the factor loadings introduces the hyperpriors:
  $\gamma_j \sim \text{Exp}(\lambda_j)$. Feature grouping is then included by
  parametrising $\forall j \in \mathcal{G}_g: \lambda_j = \lambda_g$, and
  estimating the $\lambda_g$ with empirical Bayes. In general, 
  such Gaussian scale mixture extensions of
  the $\bar{\mathbf{B}}$ prior require the addition of one or more extra 
  layers to the prior and one or more extra variational parameters to update 
  during estimation. Some existing examples of sparse Bayesian
  factor models are \cite{ferrari_bayesian_2020} and
  \cite{carvalho_high-dimensional_2008}. Sparse factor models often simplify
  the latent dimension estimation. In any case, latent dimension estimation is 
  a topic that deserves more 
  attention. Here, estimation is via a simple Kaiser criterion. More elaborate
  methods are available in literature \cite[see, e.g.,][]{auerswald_how_2019}.
	
	Lastly, we give some indication of computational times.
	The proposed factor regression approaches are slower to estimate compared to
	the other methods. Model estimation times for the influenza application are: 
	0.87, 0.16, 5.27, and 
	36.46 seconds, for the ridge, lasso,
	FMradio, and Bayesian factor regression models, respectively. 
	For the oral cancer metastasis appication we have
	2.76, 0.98 and 54.82 
	seconds for the ridge, lasso and FMradio, and
	124.68 and 245.18 
	minutes for the variational and empirical Bayesian models.
	Especially in the second application, the estimation is considerably slower.
	However, we argue that these times are still manageable and much faster
	than traditional MCMC estimation times.
	
	\bibliographystyle{author_short3} 
	\bibliography{refs}

	\newpage
	
  \title{Supplementary Material to: `Semi-supervised empirical Bayes 
  group-regularized factor regression'}
  
  \maketitle

	\section{Introduction}
	This document contains Supplementary Material (SM) to the Main Document (MD)
	titled `Semi-supervised empirical Bayes group-regularized factor regression'.
	
	\section{Model identifiability and determinancy}
	\label{sec:modelunidentifiability}
	Two issues arise from MD model (\ref{eq:model2}): (i) 
	Rotational unidentifiability
	and (ii) indeterminancy in the latent factors. Issue (i) becomes apparent if
	we multiply the loadings with an orthogonal matrix $\mathbf{H}$ and consider 
	the implied covariance matrix:
	\begin{equation}\label{eq:rotation}
	\mathbb{C}\text{ov}(\bar{\mathbf{x}}) = \bar{\mathbf{B}}^{\text{T}}
	\bar{\mathbf{B}} + \bar{\bm{\Psi}} = 
	(\mathbf{H}\bar{\mathbf{B}})^{\text{T}}(\mathbf{H}\bar{\mathbf{B}}) + 
	\bar{\bm{\Psi}}.
	\end{equation}
	From (\ref{eq:rotation}) we see that any arbitrary orthogonal transformation
	yields the same covariance. This indeterminancy is usually fixed by 
	restricting
	$\bar{\mathbf{B}} \bar{\bm{\Psi}}^{-1} \bar{\mathbf{B}}^{\text{T}}$ to a  
	diagonal matrix during maximum likelihood
	estimation. This restriction has no apparent interpretation, so to increase
	the interpretability of the model, a \textit{post hoc} rotation of the 
	loadings is often desirable. A popular choice is the orthogonal Varimax 
	rotation, 
	which maximises the sum of the variances of the squared loadings. 
	Varimax rotation often leads to approximately sparse representations of 
	features that benefit interpretability.
	
	Issue (ii), indeterminancy of latent factors, stems from the initial 
	postulate $\mathbb{E}(\bm{\lambda}\bm{\lambda}^{\text{T}})=\mathbf{I}_d$. 
	The predicted scores $\hat{\bm{\lambda}} = 
	\mathbb{E}(\bm{\lambda} | \bar{\mathbf{x}})$
	do not necessarily adhere to this postulate:
	$$
	\mathbb{E}(\hat{\bm{\lambda}}\hat{\bm{\lambda}}^{\text{T}}) = 
	\bar{\mathbf{B}} ( \bar{\mathbf{B}}^{\text{T}} \bar{\mathbf{B}} + 
	\bar{\bm{\Psi}})^{-1}
	\bar{\mathbf{B}}^{\text{T}} \neq \mathbf{I}_d.
	$$
	However, we can generate a random variable $\mathbf{s}$, uncorrelated to 
	$\bar{\mathbf{x}}$, and with expectation $\mathbf{0}_d$ and variance 
	$\mathbf{I}_d$
	to construct scores:
	$$
	\hat{\bm{\lambda}} = 
	\bar{\mathbf{B}} ( \bar{\mathbf{B}}^{\text{T}} \bar{\mathbf{B}} + 
	\bar{\bm{\Psi}})^{-1}
	\bar{\mathbf{B}}^{\text{T}} + \[\mathbf{I}_d - 
	\bar{\mathbf{B}} (\bar{\mathbf{B}}^{\text{T}}\bar{\mathbf{B}} + 
	\bar{\bm{\Psi}})^{-1} \bar{\mathbf{B}}^{\text{T}}\]^{1/2}\mathbf{s},
	$$
	that do adhere to the postulate. Unfortunately, there are infinitely many 
	choices
  of $\mathbf{s}$, leading to an indeterminancy in the latent factors. A 
  simple fix to
  this indeterminancy is setting
  $\bar{\bm{\Psi}} = n^{-1}[(\bar{\mathbf{X}}^{\text{T}} 
  \bar{\mathbf{X}})^{-1} \circ \mathbf{I}_p]^{-1}$. 
  
	\section{Maximum likelihood estimation}\label{sec:mlestimation}
	\subsection{Maximum likelihood}\label{sec:maximumlikelihood}
	In the low-dimensional setting ($\bar{p} < n$), loadings $\bar{\mathbf{B}}$ 
	and variances
	$\bar{\psi}_1, \dots, \bar{\psi}_{\bar{p}}$ in MD model (\ref{eq:model2}) 
	may be estimated by 
	maximum likelihood. If we denote the parameters
	by $\theta=\{ \bar{\mathbf{B}}, \bar{\psi}_1, \dots, \bar{\psi}_{\bar{p}} \}$, 
	the maximum 
	likelihood estimate is given by \cite[]{mardia_multivariate_1979}:
	\begin{equation}\label{eq:mle}
	  \hat{\theta} = \underset{\theta}{\argmax} \log|\bar{\bm{\Sigma}}^{-1}| -
	  \text{tr}\( \bar{\bm{\Sigma}}^{-1} \bar{\mathbf{S}}\),
	\end{equation}
	where $\bar{\bm{\Sigma}}=\bar{\mathbf{B}}^{\text{T}}\bar{\mathbf{B}} + 
	\bar{\bm{\Psi}}$, and
	$\bar{\mathbf{S}} = n^{-1} \bar{\mathbf{X}}^{\text{T}}\bar{\mathbf{X}}$
	is the empirical covariance matrix of $\bar{\mathbf{X}}$. The maximum 
	likelihood estimation is 
	implemented in the base \texttt{R} 
	package \texttt{stats} as the \texttt{factanal} function.
	
	\subsection{Penalised maximum likelihood}
	\label{sec:penalizedmaximumlikelihood}
	In high dimensional settings, i.e., $\bar{p} > n$, the unique (up to 
	rotation) maximum likelihood estimate of MD model (\ref{eq:model2}) does 
	not exist. One solution is to penalise the
	likelihood \cite[]{van_wieringen_ridge_2016}:
	\begin{equation}\label{eq:penalisedmle}
	  \hat{\theta} = \underset{\theta}{\argmax} \log|\bar{\bm{\Sigma}}^{-1} | - 
	  (1 - \gamma) \text{tr}\( \bar{\bm{\Sigma}}^{-1} 
	  \bar{\mathbf{S}}\) - 
	  \gamma \text{tr} \(\bar{\bm{\Sigma}}^{-1} \bm{\Gamma}\),
	\end{equation}
	where $\gamma \in \(0, 1 \]$ is a penalty parameter that determines the 
	amount of penalisation and $\bm{\Gamma}$ is a positive definite shrinkage 
	target matrix, usually taken to be diagonal, or even the identity. Computing
	(\ref{eq:penalisedmle}) is easily done via \texttt{factanal} in 
	\texttt{R}, where the empirical covariance is replaced with its shrunken
	estimate $(1 - \gamma) \bar{\mathbf{S}} + \gamma \bm{\Gamma}$. 
	Common choices for target $\bm{\Gamma}$ are \cite[]{van_wieringen_ridge_2016}
	$\bm{\Gamma}=\mathbf{I}_p$ and $\bm{\Gamma}$ diagonal with 
	$\bm{\Gamma}_{jj}=\bar{\mathbf{S}}_{jj}$. A convenient and data-drive method 
	to pick penalty parameter $\gamma$ is
	$k$-fold cross validation, as described in \cite{peeters_stable_2019}. 
	
	The attentive reader may have noticed that the penalty in 
	(\ref{eq:penalisedmle}) is not a proper penalty. The added penalty term is
	$\gamma \text{tr}[ \bar{\bm{\Sigma}}^{-1} (\bm{\Gamma}- \bar{\mathbf{S}}) ]$, 
	which does not necessarily penalise the likelihood if 
	$\bm{\Gamma}-\bar{\mathbf{S}} > 0$ 
	somewhere. It is however, a ``penalty'' 
	with good empirical performance. In addition, it automatically models
	a correlation matrix if both the empirical covariance and shrinkage targets
	are proper correlation matrices.
	
	\subsection{Unlabeled features}\label{sec:unlabeled}
	To incorporate the 
	observed, unlabeled features $\mathbf{x}_i$, for $i=n + 1, \dots, n + m$, we 
	treat the unobserved, corresponding responses $z_i$, for $i=n + 1, \dots, 
	n + m$, as missing and employ an EM algorithm to incorporate them into 
	the likelihood maximisation. Writing 
	$\mathbf{z} = \begin{bmatrix} z_{n+1} & \cdots & z_{n+m}
	\end{bmatrix}^{\text{T}}$ for the unobserved responses and 
	$\theta^{(k)}=\{ \bar{\mathbf{B}}^{(k)}, \bar{\psi}_1^{(k)}, \dots,
	\bar{\psi}_{\bar{p}}^{(k)}\}$
	for the current parameter estimates, we have respective E- and M-steps:
	\begin{subequations}\label{eq:em}
	  \begin{align}
	    Q(\theta | \theta^{(k)}) & = \mathbb{E}_{\mathbf{z} | \mathbf{y}, 
	    \mathbf{X}} \[\log p \(\mathbf{z}, \mathbf{y}, \mathbf{X} \) | 
	    \theta^{(k)} \], \label{eq:expectedloglikelihood}\\
	    \theta^{(k + 1)} & = \underset{\theta}{\argmax} Q(\theta | \theta^{(k)}),
	  \end{align}
	\end{subequations}
	which we iteratively apply until convergence of the expected log likelihood
	(\ref{eq:expectedloglikelihood}). 
	
	The E-step in (\ref{eq:expectedloglikelihood}) is:
	\begin{subequations}\label{eq:estep}
	  \begin{align*}
	    \mathbb{E}_{\mathbf{z} | \mathbf{y}, \mathbf{X}} 
	    \[\log p \(\mathbf{z}, \mathbf{y}, \mathbf{X} \) | \theta^{(k)} \] \propto
	    \sum_{i=1}^n \log p(y_i, \mathbf{x}_i | \theta^{(k)} ) +
	    \sum_{i=n + 1}^{n + m} \mathbb{E}_{z_i | \mathbf{x}_i} 
	    \[ \log p(z_i, \mathbf{x}_i | \theta^{(k)} ) \].
	  \end{align*}
	\end{subequations}
	The first term is just the regular likelihood that is maximised in 
	(\ref{eq:mle}). The second term may be written as:
	\begin{align*}
	  \sum_{i=n + 1}^{n + m} \mathbb{E}_{z_i | \mathbf{x}_i} 
	  \[ \log p(z_i, \mathbf{x}_i | \theta^{(k)} ) \] & \propto 
	  \log |\bar{\bm{\Sigma}}| - m^{-1} \sum_{i=n + 1}^{n + m} 
	  \mathbb{E}_{z_i | \mathbf{x}_i} 
	  \( \begin{bmatrix} \mathbf{x}_i \tr & z_i \end{bmatrix} 
	  \bar{\bm{\Sigma}}^{-1} \begin{bmatrix} \mathbf{x}_i \tr & z_i 
	  \end{bmatrix}^{\text{T}}\) \\
	  & = \log |\bar{\bm{\Sigma}}| - 
	  \text{tr} \[ \bar{\bm{\Sigma}}^{-1}
	  m^{-1} \sum_{i=n + 1}^{n + m} \mathbb{E}_{z_i | \mathbf{x}_i} 
	  \( \begin{bmatrix} \mathbf{x}_i \tr & z_i 
	  \end{bmatrix}^{\text{T}} \begin{bmatrix} \mathbf{x}_i \tr & z_i 
	  \end{bmatrix}\) \] \\
	  & = \log |\bar{\bm{\Sigma}}| - \text{tr} \( \bar{\bm{\Sigma}}^{-1}
	  \tilde{\mathbf{S}}_{-n}\),
	\end{align*}
	where we have written:
	\begin{align*}
	  \tilde{\mathbf{S}}_{-n} & = m^{-1} \sum_{i=n + 1}^{n + m} 
	  \mathbb{E}_{z_i | \mathbf{x}_i} \( \begin{bmatrix} \mathbf{x}_i \tr & z_i 
	  \end{bmatrix}^{\text{T}} \begin{bmatrix} \mathbf{x}_i \tr & z_i 
	  \end{bmatrix}\) \\
	  & = m^{-1} \sum_{i=n + 1}^{n + m} \[ \mathbb{E}_{z_i | \mathbf{x}_i} \(
	  \begin{bmatrix} \mathbf{x}_i \tr & z_i \end{bmatrix}^{\text{T}} \)
	  \mathbb{E}_{z_i | \mathbf{x}_i} \(
	  \begin{bmatrix} \mathbf{x}_i \tr & z_i \end{bmatrix} \) +
	  \mathbb{V}_{z_i | \mathbf{x}_i} \( 
	  \begin{bmatrix} \mathbf{x}_i \tr & z_i \end{bmatrix}^{\text{T}} \) \] \\
	  & = m^{-1} \( \tilde{\mathbf{X}}_{-n}^{\text{T}} \tilde{\mathbf{X}}_{-n} +
	  \begin{bmatrix} \mathbf{0}_{p \times p} & \mathbf{0}_{p \times 1} \\
	  \mathbf{0}_{1 \times p} & m \mathbb{V}_{z | \mathbf{x}} 
	  (z | \theta^{(k)}) \end{bmatrix} \),
	\end{align*}
	with 
	\begin{equation}\label{eq:expectation}
	\tilde{\mathbf{X}}_{-n}^{\text{T}} = \begin{bmatrix} 
	\mathbf{x}_{n+1}^{\text{T}} & \mathbb{E}_{z_{n+1} | \mathbf{x}_{n+1}}
	(z_{n+1} | \theta^{(k)}) \\
	\vdots & \vdots \\
	\mathbf{x}_{n+m}^{\text{T}} & \mathbb{E}_{z_{n+m} | \mathbf{x}_{n+m}}
	(z_{n+m} | \theta^{(k)}) \\
	\end{bmatrix}.
	\end{equation}
	
	Now if we combine this term with the regular likelihood term we have:
	\begin{equation}\label{eq:expectedloglikelihood2}
	  \mathbb{E}_{\mathbf{z} | \mathbf{y}, \mathbf{X}} 
	  \[\log p \(\mathbf{z}, \mathbf{y}, \mathbf{X} \) | \theta^{(k)} \] \propto
	  \log|\bar{\bm{\Sigma}}^{-1}| - \text{tr}\( \bar{\bm{\Sigma}}^{-1}
	  \tilde{\mathbf{S}}\),
	\end{equation}
	where
	\begin{equation}\label{eq:variance}
	\tilde{\mathbf{S}} = (n + m)^{-1}\(\tilde{\mathbf{X}}^{\text{T}}
	\tilde{\mathbf{X}} + \begin{bmatrix} 
	\mathbf{0}_{p \times p} & \mathbf{0}_{p \times 1} \\
	\mathbf{0}_{1 \times p} & m \mathbb{V}_{z | \mathbf{x}} 
	(z | \theta^{(k)})\end{bmatrix}\),
	\end{equation}
	and
	\begin{equation}
	\tilde{\mathbf{X}} = \begin{bmatrix} 
	\bar{\mathbf{X}} \\
	\tilde{\mathbf{X}}_{-n}
	\end{bmatrix}.
	\end{equation}
	The expectations and variance in (\ref{eq:expectation}) and 
	(\ref{eq:variance}) are easily derived from the well-known relation between
	the joint multivariate normal distribution in (\ref{eq:model2}) and the
	corresponding conditional distributions:
	\begin{subequations}
	  \begin{align*}
	  \mathbb{E}_{z_i | \mathbf{x}_i} 
	    (z_i | \theta^{(k)}) & = \mathbf{x}_i^{\text{T}}
	    \[(\mathbf{B}^{(k)})^{\text{T}}\mathbf{B}^{(k)}
	    + \bm{\Psi}^{(k)}\]^{-1} (\mathbf{B}^{(k)})^{\text{T}}\bm{\beta}^{(k)}, \\
	    \mathbb{V}_{z | \mathbf{x}} (z | \theta^{(k)}) & = 
	    (\bm{\beta}^{(k)})^{\text{T}} \left\{ \mathbf{I}_d - 
	    \mathbf{B}^{(k)} \[(\mathbf{B}^{(k)})^{\text{T}}\mathbf{B}^{(k)}
	    + \bm{\Psi}^{(k)}\]^{-1} (\mathbf{B}^{(k)})^{\text{T}} \right\} 
	    \bm{\beta}^{(k)}  + (\sigma^2)^{(k)}.
	  \end{align*}
	\end{subequations}
	The M-step in (\ref{eq:expectedloglikelihood2}) is easily computed as before 
	in (\ref{eq:mle}), using the \texttt{factanal} function in \texttt{R} with the
	augmented empirical covariance $\tilde{\mathbf{S}}$ instead of 
	$\bar{\mathbf{S}}$. The above is easily
	generalised to include data with arbitrary missingness patterns.
	
	\section{Bayesian inference}
	In this Section it is useful to note that the likelihood MD 
	(\ref{eq:model2}) of i.i.d. Gaussian
	data 
	$\bar{\mathbf{X}} = \begin{bmatrix} \bar{\mathbf{x}}_1 & \cdots & 
	\bar{\mathbf{x}}_n 
	\end{bmatrix}^{\text{T}}$ may be rewritten as a product over densities of
	columns $\bar{\mathbf{x}}_{\bar{j}}$, $\bar{j}=1, \dots, \bar{p}$, instead of 
	the observations $\bar{\mathbf{x}}_i$, $i=1, \dots, n$, where we abuse 
	notation to indicate column (and thus variable) $\bar{j}$ of 
	$\bar{\mathbf{X}}$ by $\bar{\mathbf{x}}_{\bar{j}}$ and row 
	(and thus observation) $i$ by $\bar{\mathbf{x}}_i$:
	\begin{align}\label{eq:likelihood2}
	  \prod_{i=1}^n p(\bar{\mathbf{x}}_i | \bm{\lambda}_i, \bar{\mathbf{B}}, 
	    \bar{\psi}_1, \dots, \bar{\psi}_{\bar{p}}) = 
	    \prod_{\bar{j}=1}^{\bar{p}} p(\bar{\mathbf{x}}_{\bar{j}} | \bm{\Lambda}, 
	    \bar{\mathbf{b}}_{\bar{j}}, \bar{\psi}_{\bar{j}}),
	\end{align}
	with $\bm{\Lambda} = \begin{bmatrix} \bm{\lambda}_1 & \cdots & \bm{\lambda}_n
	\end{bmatrix}^{\text{T}}$ and 
	$$
	p(\bar{\mathbf{x}}_{\bar{j}} | \bm{\Lambda}, \bar{\mathbf{b}}_{\bar{j}}, 
	  \bar{\psi}_{\bar{j}}) \overset{D}{=}
	  \mathcal{N}_n \( \bm{\Lambda} \bar{\mathbf{b}}_{\bar{j}}, \psi_{\bar{j}} 
	  \mathbf{I}_n \).
	$$
	
	\subsection{Gibbs sampling}\label{sec:gibbssampling}
	This Section derives the full conditionals of the Bayesian model in 
	(\ref{eq:model2}) and (\ref{eq:prior}). To that end, in this Section,
	we include the unlabeled features and unobserved outcomes into 
	$\bar{\mathbf{X}}$, i.e., 
  $$
  \bar{\mathbf{X}}_{(n+m) \times \bar{p}} = \begin{bmatrix}
    \multicolumn{1}{c}{\multirow{2}{*}{$\mathbf{X}$}} & \mathbf{y} \\
    \multicolumn{1}{c}{} & \mathbf{z}
  \end{bmatrix}.
  $$
  In addition, we slightly abuse notation, write $\bar{\mathbf{x}}_i$ and
  $\bar{\mathbf{x}}_j$ as the $i$th row and $j$th column of $\bar{\mathbf{X}}$,
  respectively. Then, the full conditional for $\bm{\Lambda}$ is:
	\begin{align*}
	  p(\bm{\Lambda} | \bar{\mathbf{X}}, \bar{\mathbf{B}}, 
  	  \bar{\psi}_1, \dots \bar{\psi}_{\bar{p}}) & \propto \prod_{i=1}^{n + m} 
  	  p(\bar{\mathbf{x}}_i | \bm{\lambda}_i, \bar{\mathbf{B}}, \bar{\psi}_1, 
  	  \dots \bar{\psi}_{\bar{p}}) \prod_{i=1}^{n + m} p(\bm{\lambda}_i) \\
    & \propto \prod_{i=1}^{n + m} \exp \[ -\frac{1}{2} 
      (\bar{\mathbf{x}}_i - \bar{\mathbf{B}}^{\text{T}} 
      \bm{\lambda}_i)^{\text{T}} 
      \bar{\bm{\Psi}}^{-1} (\bar{\mathbf{x}}_i - \bar{\mathbf{B}}^{\text{T}} 
      \bm{\lambda}_i)
      - \frac{1}{2}\bm{\lambda}_i^{\text{T}} \bm{\lambda}_i \] \\
    & \propto \prod_{i=1}^{n + m} \exp \bigg\{ -\frac{1}{2} 
      \[ \bm{\lambda}_i - (\bar{\mathbf{B}} \bar{\bm{\Psi}}^{-1} 
      \bar{\mathbf{B}}^{\text{T}} + \mathbf{I}_d)^{-1} \bar{\mathbf{B}} 
      \bar{\bm{\Psi}}^{-1} \bar{\mathbf{x}}_i\]^{\text{T}} \\
    & \quad \quad \quad (\bar{\mathbf{B}} \bar{\bm{\Psi}}^{-1} 
      \bar{\mathbf{B}}^{\text{T}} + \mathbf{I}_d)
      \[ \bm{\lambda}_i - (\bar{\mathbf{B}} \bar{\bm{\Psi}}^{-1} 
      \bar{\mathbf{B}}^{\text{T}} + \mathbf{I}_d)^{-1} \bar{\mathbf{B}} 
      \bar{\bm{\Psi}}^{-1} \bar{\mathbf{x}}_i\] \bigg\},
	\end{align*}
	which allows us to write:
	\begin{align}
	  \bm{\Lambda} | \bar{\mathbf{X}}, \bar{\mathbf{B}}, 
	  \bar{\psi}_1, \dots \bar{\psi}_{\bar{p}} & \sim \prod_{i=1}^{n + m} 
	  \mathcal{N}_d
	  \((\bar{\mathbf{B}} \bar{\bm{\Psi}}^{-1}\bar{\mathbf{B}}^{\text{T}} +
	  \mathbf{I}_d)^{-1}\bar{\mathbf{B}}\bar{\bm{\Psi}}^{-1} \bar{\mathbf{x}}_i,
	  (\bar{\mathbf{B}} \bar{\bm{\Psi}}^{-1}\bar{\mathbf{B}}^{\text{T}} +
	  \mathbf{I}_d)^{-1}\). \label{eq:conditionalLambda}
  \end{align}
  
  Next, we consider 
  \begin{align*}
	  p(\bar{\mathbf{B}} | \bar{\mathbf{X}}, \bm{\Lambda}, 
  	  \bar{\psi}_1, \dots \bar{\psi}_{\bar{p}}) & \propto \prod_{i=1}^{n + m} 
  	  p(\bar{\mathbf{x}}_i | \bm{\lambda}_i, \bar{\mathbf{B}}, \bar{\psi}_1, 
  	  \dots \bar{\psi}_{\bar{p}}) \prod_{\bar{j}=1}^{\bar{p}} 
  	  p(\bar{\mathbf{b}}_{\bar{j}} | \bar{\psi}_{\bar{j}}) \\
  	& = \prod_{\bar{j}=1}^{\bar{p}} p(\bar{\mathbf{x}}_{\bar{j}} | 
  	  \bm{\Lambda}, \bar{\mathbf{b}}_{\bar{j}}, \bar{\psi}_{\bar{j}})
  	  p(\bar{\mathbf{b}}_{\bar{j}} | \bar{\psi}_{\bar{j}}) \\
  	& \propto \prod_{\bar{j}=1}^{\bar{p}} 
  	  \exp \[ \frac{\bar{\psi}_{\bar{j}}^{-1}}{2} 
  	  (\bar{\mathbf{x}}_{\bar{j}} - \bm{\Lambda} 
  	  \bar{\mathbf{b}}_{\bar{j}})^{\text{T}} (\bar{\mathbf{x}}_{\bar{j}} - 
  	  \bm{\Lambda} \bar{\mathbf{b}}_{\bar{j}}) - 
  	  \frac{\gamma_{\bar{j}}^{-1} \bar{\psi}_{\bar{j}}^{-1}}{2}
  	  \bar{\mathbf{b}}_{\bar{j}}^{\text{T}} \bar{\mathbf{b}}_{\bar{j}} \] \\
    & \propto \prod_{\bar{j}=1}^{\bar{p}} \exp \bigg\{ -\frac{1}{2} 
      \[ \bar{\mathbf{b}}_{\bar{j}} - ( \bm{\Lambda}^{\text{T}}\bm{\Lambda} + 
  	  \gamma^{-1}_{\bar{j}}\mathbf{I}_d)^{-1}
  	  \bm{\Lambda}^{\text{T}} \bar{\mathbf{x}}_{\bar{j}}\]^{\text{T}} \\
    & \quad \quad \quad 
      \bar{\psi}_{\bar{j}}^{-1} (\bm{\Lambda}^{\text{T}}\bm{\Lambda} + 
  	  \gamma^{-1}_{\bar{j}}\mathbf{I}_d) \[ \bar{\mathbf{b}}_{\bar{j}} - 
  	  ( \bm{\Lambda}^{\text{T}}\bm{\Lambda} + 
  	  \gamma^{-1}_{\bar{j}}\mathbf{I}_d)^{-1}
  	  \bm{\Lambda}^{\text{T}} \bar{\mathbf{x}}_{\bar{j}}\] \bigg\},
	\end{align*}
	which gives
	\begin{align}
	\bar{\mathbf{B}} | \bar{\mathbf{X}}, \bm{\Lambda}, 
	  \bar{\psi}_1, \dots \bar{\psi}_{\bar{p}} \sim \prod_{\bar{j}=1}^{\bar{p}}
  	\mathcal{N}_{d} \( ( \bm{\Lambda}^{\text{T}}\bm{\Lambda} + 
  	\gamma^{-1}_{\bar{j}}\mathbf{I}_d)^{-1}
  	\bm{\Lambda}^{\text{T}} \bar{\mathbf{x}}_{\bar{j}},
  	\bar{\psi}_{\bar{j}} (\bm{\Lambda}^{\text{T}}\bm{\Lambda} + 
  	  \gamma^{-1}_{\bar{j}}\mathbf{I}_d)^{-1}\). \label{eq:conditionalB}
	\end{align}
	
	For the $\bar{\psi}_{\bar{j}}$, we derive:
	\begin{align*}
	  p(\bar{\psi}_1, \dots \bar{\psi}_{\bar{p}} | \bar{\mathbf{X}}, \bm{\Lambda}, 
  	  \bar{\mathbf{B}}) & \propto \prod_{i=1}^{n + m} 
  	  p(\bar{\mathbf{x}}_i | \bm{\lambda}_i, \bar{\mathbf{B}}, \bar{\psi}_1, 
  	  \dots \bar{\psi}_{\bar{p}}) \prod_{\bar{j}=1}^{\bar{p}} 
  	  p(\bar{\mathbf{b}}_{\bar{j}} | \bar{\psi}_{\bar{j}})
  	  p(\bar{\psi}_{\bar{j}}) \\
  	& = \prod_{\bar{j}=1}^{\bar{p}} p(\bar{\mathbf{x}}_{\bar{j}} | 
  	  \bm{\Lambda}, \bar{\mathbf{b}}_{\bar{j}}, \bar{\psi}_{\bar{j}})
  	  p(\bar{\mathbf{b}}_{\bar{j}} | \bar{\psi}_{\bar{j}})
  	  p(\bar{\psi}_{\bar{j}}) \\
  	& \propto \prod_{\bar{j}=1}^{\bar{p}} 
  	  \bar{\psi}_{\bar{j}}^{-(\frac{n + m + d}{2} + \kappa_{\bar{j}}) - 1} \\
  	& \quad \quad \quad \exp \left\{ -\bar{\psi}_{\bar{j}}^{-1} \[\frac{1}{2} 
  	   (\bar{\mathbf{x}}_{\bar{j}} - 
  	  \bm{\Lambda}\bar{\mathbf{b}}_{\bar{j}})^{\text{T}}
  	  (\bar{\mathbf{x}}_{\bar{j}} - 
  	  \bm{\Lambda}\bar{\mathbf{b}}_{\bar{j}}) + \frac{\gamma^{-1}_{\bar{j}}}{2}
  	  \bar{\mathbf{b}}_{\bar{j}}^{\text{T}}\bar{\mathbf{b}}_{\bar{j}} + 
  	  \nu_{\bar{j}}\] \right\},
	\end{align*}
	to arrive at
	\begin{align}
	\bar{\psi}_1, \dots \bar{\psi}_{\bar{p}} | \bar{\mathbf{X}}, \bm{\Lambda}, 
	  \bar{\mathbf{B}} \sim \prod_{\bar{j}=1}^{\bar{p}} 
	  \Gamma^{-1}\(\frac{n + m + d}{2} + \kappa_{\bar{j}},
	  \frac{1}{2}(\bar{\mathbf{x}}_{\bar{j}} - 
	  \bm{\Lambda}\bar{\mathbf{b}}_{\bar{j}})^{\text{T}}
	  (\bar{\mathbf{x}}_{\bar{j}} - 
	  \bm{\Lambda}\bar{\mathbf{b}}_{\bar{j}}) + 
	  \frac{\gamma_{\bar{j}}^{-1}}{2}\bar{\mathbf{b}}_{\bar{j}}^{\text{T}}
	  \bar{\mathbf{b}}_{\bar{j}} + \nu_j \). \label{eq:conditionalPsi} 
	\end{align}
	
	Given the latent variables and parameters, $z$ is independent of $\mathbf{x}$,
	so the full conditional for the missing outcomes is equal to the likelihood:
	\begin{align}\label{eq:conditionalLabels}
	\mathbf{z} | \mathbf{X}, \bm{\Lambda}, \bar{\mathbf{B}},
	  \bar{\psi}_1, \dots \bar{\psi}_{\bar{p}} \sim \prod_{i=n + 1}^{n + m} 
	  \mathcal{N} (\bar{\mathbf{b}}_{\bar{p}}^{\text{T}} \bm{\lambda}_i, 
	  \bar{\psi}_{\bar{p}}).
	\end{align}
  Using these full conditionals, samples from the posterior may be generated 
  through a straightforward Gibbs sampling scheme.

  \subsection{Variational evidence lower bound}
  Variational parameters are generally updated until convergence of
  the evidence lower bound. Here we describe a general case of our
  evidence lower bound, in which missingness
  may be occur in all all variables (i.e., both response and features).
  The number of missing values for feature $\bar{j}$ is indicated with 
  $m_{\bar{j}}$.
  Let $\tau_{\bar{j}} = (n/2 + d/2 + \kappa_{\bar{j}})/(2\zeta_{\bar{j}})$, then:
  \begin{align*}
    \text{ELBO}= & -\frac{np - \sum_{{\bar{j}}=1}^{\bar{p}} m_{\bar{j}}}{2} 
      \log 2 \pi + 
      \frac{nd + pn + dp + \sum_{{\bar{j}}=1}^{\bar{p}} m_{\bar{j}}}{2} \\
    & + \sum_{{\bar{j}}=1}^{{\bar{p}}} \left[ \log \Gamma \( \frac{n + d}{2} +
      \kappa_{\bar{j}} \) - \log \Gamma(\kappa_{\bar{j}}) - 
      \frac{d}{2} \psi \(\frac{n + d}{2} + \kappa_{\bar{j}}\)
      + \frac{d}{2}\log \gamma_{\bar{j}} + \kappa_{\bar{j}} 
      \(1 + \log \nu_{\bar{j}} \) \right] \\
    & + \sum_{{\bar{j}}=1}^{{\bar{p}}} \left[\frac{m_{\bar{j}}}{2} 
      \log \chi_{\bar{j}} - \tau_{\bar{j}} m_{\bar{j}} \chi_{\bar{j}} -
      \left( \frac{n - d}{2} + \kappa_{\bar{j}} \right) \log \zeta_{\bar{j}} 
      - \frac{1}{2} \nu_{\bar{j}} \tau_{\bar{j}} \right] \\
    & + \sum_{{\bar{j}}=1}^{\bar{p}} \[\frac{1}{2} 
      \log |\bm{\Omega}_{\bar{j}}| -
      \tau_{\bar{j}} \gamma_{\bar{j}}^{-1} \text{tr}\(\bm{\Omega}_{\bar{j}}\) - 
      n \tau_{\bar{j}} \text{tr}\(\bm{\Xi}\bm{\Omega}_{\bar{j}}\) - 
      \tau_{\bar{j}} \text{tr}\(\bm{\Phi}^{\text{T}}\bm{\Phi}
      \bm{\Omega}_{\bar{j}}\)\] \\
    & + 2 \text{tr} \[\text{diag} \( \tau_{\bar{j}} \) 
      \tilde{\mathbf{X}}^{\text{T}} \bm{\Phi} \mathbf{M}\]
      - n\text{tr} \[\text{diag} \( \tau_{\bar{j}} \) 
      \mathbf{M}^{\text{T}}\bm{\Xi}\mathbf{M}\] -
      \text{tr} \[\text{diag} \( \tau_j \gamma_{\bar{j}}^{-1} \) 
      \mathbf{M}^{\text{T}}\mathbf{M}\] \\
    & - \text{tr} \[\text{diag} \( \tau_{\bar{j}} \) 
      \mathbf{M}^{\text{T}}\bm{\Phi}^{\text{T}}\bm{\Phi}\mathbf{M}\] -
      \text{tr} \[\text{diag} \( \tau_{\bar{j}} \) 
      \tilde{\mathbf{X}}^{\text{T}}\tilde{\mathbf{X}}\] \\  
    & + \frac{n}{2} \log |\bm{\Xi}| - \frac{n}{2} \text{tr}( \bm{\Xi}) -
      \frac{1}{2}\text{tr} \(\bm{\Phi}^{\text{T}} \bm{\Phi}\),
  \end{align*}
  where $\psi(x)$ denotes the digamma function and 
  $\mathbf{M}=\begin{bmatrix} \bm{\mu}_1 & \cdots & \bm{\mu}_{{\bar{p}}} 
  \end{bmatrix}$.
  
  % \subsection{Empirical Bayes}\label{sec:empiricalbayes}
  % To estimate $\bm{\gamma} = \begin{bmatrix} \gamma_1 & \cdots & \gamma_G 
  % \end{bmatrix}^{\text{T}}$ by empirical Bayes, we write
  % $\theta = \{ \bar{\mathbf{B}}, \bar{\psi}_1, \dots, \bar{\psi}_{p+1}\}$ and 
  % apply the following EM steps
  % until convergence \cite[see][]{casella_empirical_2001}:
  % $$
  % \bm{\gamma}^{(k + 1)} = \underset{\bm{\gamma}}{\argmax} 
  % \mathbb{E}_{ \theta | \mathbf{y}} \[\log p(\bar{\mathbf{B}} | 
  % \bar{\psi}_1, \dots, \bar{\psi}_{p+1} ) | 
  % \bm{\gamma}^{(k)} \].
  % $$
  % The difficult expectation here is replaced with an approximate variational 
  % version, such that we have:
  % $$
  % \bm{\gamma}^{(k + 1)} = \underset{\bm{\gamma}}{\argmax}
  % - \frac{1}{2} \sum_{g=1}^G \gamma_g^{-1} 
  % \sum_{j \in \mathcal{G}_g} \mathbb{E}(\bar{\psi}_j^{-1}) 
  % \left\{ \text{tr} \[ 
  % \mathbb{V}(\bar{\mathbf{b}}_j)\] + \mathbb{E}(\bar{\mathbf{b}}_j^{\text{T}})
  % \mathbb{E}(\bar{\mathbf{b}}_j) \right\} - \frac{d}{2} 
  % \sum_{g=1}^{G} |\mathcal{G}_g| \log \gamma_{G}
  % $$
  % as our iterations. This gives empirical Bayes updates:
  % Bayes updates:
  % $$
  % \gamma_g^{(k + 1)} = \frac{\sum_{j \in \mathcal{G}_g} 
  % \mathbb{E}(\bar{\psi}_j^{-1}) \left\{ \text{tr} \[ 
  % \mathbb{V}(\bar{\mathbf{b}}_j)\] + \mathbb{E}(\bar{\mathbf{b}}_j^{\text{T}})
  % \mathbb{E}(\bar{\mathbf{b}}_j) \right\}}{|\mathcal{G}_g| d}.
  % $$
  
  \subsection{Posterior expectation}\label{sec:posteriorexpectation}
	In the (variational) Bayesian model, the prediction rule 
	$\mathbb{E}(\tilde{y} | \tilde{\mathbf{x}})$ is not available in closed-form. 
	In the MD, we approximate it with Monte Carlo simulations from the posterior.
	This is generally fast, because it requires sample from multivariate 
	Gaussian and inverse Gamma distributions, for which fast sampling algorithms
	are available.
	
	Alternatively, one may approximate the expectation with a truncated Taylor 
	series. If we denote $\theta = \{ \mathbf{B}, 
	\bm{\beta}, \bm{\psi}_1, \dots, \psi_p \}$ and $g(\theta) = 
	\tilde{\mathbf{x}}^{\text{T}}(\mathbf{B}^{\text{T}}\mathbf{B} + 
	\bm{\Psi})^{-1} \mathbf{B}^{\text{T}} \bm{\beta}$, a second order Taylor 
	approximation around the variational posterior mean of the parameters is:
	\begin{align*}
	  \mathbb{E}(\tilde{y} | \tilde{\mathbf{x}}) & = \mathbb{E}\left[g(\theta)
	    \right] \\
	  & \approx g\left[\mathbb{E}(\theta) \right] + 
	    \frac{1}{2} \text{tr} \[ \frac{\partial^2 g(\theta)}{\partial \theta^2}
	    \bigg|_{\theta=\mathbb{E}(\theta)} \mathbb{V}(\theta)\].
	\end{align*}
	The mean-field assumption and form of the variational posterior allow us to 
	write:
	\begin{align}\label{eq:taylor}
	  \mathbb{E}(\tilde{y} | \tilde{\mathbf{x}}) & \approx 
	    g\left[\mathbb{E}(\theta) \right] + 
	    \frac{1}{2} \sum_{j=1}^p \text{tr} \[ \frac{\partial^2 g(\theta)}
	    {\partial \mathbf{b}_j \partial \mathbf{b}_j^{\text{T}}}
	    \bigg|_{\mathbf{b}_j=\mathbb{E}(\mathbf{b}_j)} 
	    \mathbb{V}(\mathbf{b}_j)\] + \frac{1}{2} \sum_{j=1}^p 
	    \[ \frac{\partial^2 g(\theta)}{\partial \psi_j^2}
	    \bigg|_{\psi_j=\mathbb{E}(\psi_j)} \mathbb{V}(\psi_j)\],
	\end{align}
	where we have used that 
	$$
	\frac{\partial^2 g(\theta)} 
	{\partial \bm{\beta} \partial \bm{\beta}^{\text{T}}} = \mathbf{0}.
	$$
	The derivative in the third term of the right-hand side of (\ref{eq:taylor}) 
	is given by:
	$$
	\frac{\partial^2 g(\theta)}{\partial \psi_j^2} = 
	2 \tilde{\mathbf{x}}^{\text{T}} [(\mathbf{B}^{\text{T}}\mathbf{B} + 
	\bm{\Psi})^{-1}]_{jj} [(\mathbf{B}^{\text{T}}\mathbf{B} + 
	\bm{\Psi})^{-1}]_j^{\text{T}} [(\mathbf{B}^{\text{T}}\mathbf{B} + 
	\bm{\Psi})^{-1}]_j \mathbf{B}^{\text{T}} \bm{\beta},
	$$
	so that we can write:
	$$
	\frac{1}{2} \sum_{j=1}^p \[ \frac{\partial^2 g(\theta)}{\partial \psi_j^2}
	\bigg|_{\psi_j=\mathbb{E}(\psi_j)} \mathbb{V}(\psi_j)\] = 
	\tilde{\mathbf{x}}^{\text{T}} \mathbf{B}^{\text{T}} \bm{\beta} 
	\text{tr} \left\{ \text{diag} \[ (\mathbf{B}^{\text{T}}\mathbf{B} + 
	\bm{\Psi})^{-1} \] (\mathbf{B}^{\text{T}}\mathbf{B} + 
	\bm{\Psi})^{-2} \text{diag} [\mathbb{V}(\psi_j)]\right\}.
	$$
	The trace involving the derivatives with respect to $\mathbf{b}_j$ in
	the second term in the right-hand side (\ref{eq:taylor}) is a bit 
	more involved. We write $\mathbf{E} = (\mathbf{B}^{\text{T}}\mathbf{B} + 
	\bm{\Psi})^{-1}$ and only retain the non-zero parts when deriving
	with respect to $\mathbf{b}_j$:
	\begin{align*}
	\partial^2 g(\theta) = 2 \tilde{\mathbf{x}}^{\text{T}} \Big[ &
	\mathbf{E} (\partial \mathbf{B})^{\text{T}}
	\mathbf{B} \mathbf{E} (\partial \mathbf{B})^{\text{T}} \mathbf{B} \mathbf{E} 
	\mathbf{B}^{\text{T}}
	+ 2 \mathbf{E} \mathbf{B}^{\text{T}}
	(\partial \mathbf{B}) \mathbf{E} (\partial \mathbf{B})^{\text{T}} \mathbf{B} 
	\mathbf{E} \mathbf{B}^{\text{T}} \\
	+ & 2 \mathbf{E} (\partial \mathbf{B})^{\text{T}}
	\mathbf{B} \mathbf{E} \mathbf{B}^{\text{T}} (\partial \mathbf{B}) \mathbf{E} 
	\mathbf{B}^{\text{T}} 
	+ 2 \mathbf{E} \mathbf{B}^{\text{T}}
	(\partial \mathbf{B}) \mathbf{E} \mathbf{B}^{\text{T}} (\partial \mathbf{B}) 
	\mathbf{E} \mathbf{B}^{\text{T}} \\
	- & 2 \mathbf{E} (\partial \mathbf{B})^{\text{T}}
	(\partial \mathbf{B}) \mathbf{E} \mathbf{B}^{\text{T}}
	- 2 \mathbf{E} (\partial \mathbf{B})^{\text{T}}
	\mathbf{B} \mathbf{E} (\partial \mathbf{B})^{\text{T}}
	- 2 \mathbf{E} \mathbf{B}^{\text{T}}
	(\partial \mathbf{B}) \mathbf{E} (\partial \mathbf{B})^{\text{T}} 
	\Big] \bm{\beta}.
	\end{align*}
	Straightforward derivation of$\mathbf{B}$ with respect to $\mathbf{b}_j$ then 
	gives a very messy expression, which we do not show here.
	Plugging into (\ref{eq:taylor}), together with 
	$g[\mathbb{E}(\theta)] = 
	\tilde{\mathbf{x}}^{\text{T}}(\mathbf{B}^{\text{T}}\mathbf{B} + 
	\bm{\Psi})^{-1} \mathbf{B}^{\text{T}} 
	\bm{\beta}|_{\theta=\mathbb{E}(\theta)}$, results in a fast approximation
	method for $\mathbb{E}(\tilde{y} | \tilde{\mathbf{x}})$ that is linear
	in $\tilde{\mathbf{x}}$.
  
  \subsection{Bayesian inference for correlation matrix}\label{sec:correlation}
  This Section considers two approaches to modelling a correlation matrix
  instead of a general covariance matrix.
  
	\subsubsection{Proper correlation modelling}
	We observe $n$ observations on $p$ standardised variables $\mathbf{x}$, 
	i.e., $\sum_{i=1}^n x_{ij} = 0, n^{-1} \sum_{i=1}^n x^2_{ij} = 1$, 
	$j=1, \dots, p$. The standardisation implies that we should model
	the correlation matrix instead of the covariance matrix. 
	Write $\bm{\Psi} = \diag(\psi_j)$, for the diagonal matrix with 
	unique variances $\psi_j$, $j=1, \dots, p$ on the diagonal, 
	$\mathbf{B}_{d \times p}$ for
	the matrix of factor loadings, $\lambda$ for a vector of $d$ latent 
	variables and consider the observational factor model:
	\begin{subequations}\label{eq:cormodel}
  	\begin{align}
  	  \mathbf{x} | \bm{\lambda}, \mathbf{B}, \psi_1, \dots, \psi_p, & \sim 
  	    \mathcal{N}_p \(\mathbf{B}^{\text{T}} \bm{\lambda}, \bm{\Psi} \), \\
  	  \bm{\lambda} & \sim \mathcal{N}_d \(\mathbf{0}_{d \times 1}, 
  	    \mathbf{I}_d\).
  	\end{align}
	\end{subequations}
	Model (\ref{eq:cormodel}) induces a marginal covariance matrix:
	$$
	\mathbb{V}(\mathbf{x}) = \mathbf{B}^{\text{T}} \mathbf{B} + \bm{\Psi},
	$$
	which is a non-degenerate correlation matrix if we require
	$\forall j: \psi_j = 1 - \mathbf{b}_j^{\text{T}} \mathbf{b}_j > 0$.
	
	We now consider the Bayesian prior
	\begin{align}\label{eq:corprior}
	  \mathbf{B}, \psi_1, \dots, \psi_p & \sim \prod_{j=1}^{p}
	    \mathcal{N}_d(\mathbf{0}_{d \times 1}, \gamma_j \mathbf{I}_d) 
	    \mathbbm{1}\{ \mathbf{b}_j^{\text{T}} \mathbf{b}_j < 1 \}
	    \delta \(\psi_j - 1 + \mathbf{b}_j^{\text{T}} \mathbf{b}_j \).
	\end{align}
	This prior is a product of priors over variables $j=1, \dots, p$, with
	each variable $j$ prior itself a product of a multivariate normal for
	$\mathbf{b}_j$, truncated to a unit ball and a (degenerate) Dirac 
	distribution for $\psi_j$. Introduction of the latent variables
	$\psi_j$ results in tractable (approximate) posterior computations, as will 
	become clear later on. To see that prior
	(\ref{eq:corprior}) results in a model for a correlation matrix, we use 
	(\ref{eq:likelihood2}) to integrate
	the joint distribution of data and prior over the $\psi_j$:
	\begin{align*}
	  \int_{\psi_1} \cdots \int_{\psi_p} &
	    \prod_{i=1}^n p(\mathbf{x}_i | \bm{\lambda}_i, \mathbf{B}, \psi_1, \dots, 
	    \psi_p) p(\mathbf{B}, \psi_1, \dots, \psi_j) d\psi_1 \cdots d\psi_j \\
	  & = \prod_{j=1}^p \int_{\psi_j} 
	    p(\mathbf{x}_j | \bm{\Lambda}, \mathbf{b}_j, \psi_j) 
	    p(\mathbf{b}_j, \psi_j) d\psi_j 
	  = \prod_{j=1}^p p(\mathbf{x}_j | \bm{\Lambda}, \mathbf{b}_j) 
	    p(\mathbf{b}_j) \\
	  & = \prod_{i=1}^n p(\mathbf{x}_i | \bm{\lambda}_i, \mathbf{B})
	    p(\mathbf{B}),
	\end{align*}
	where $p(\mathbf{x}_i | \bm{\lambda}_i, \mathbf{B})$ corresponds to the
	observational model (\ref{eq:cormodel}) with 
	$\forall j: \psi_j = 1 - \mathbf{b}_j^{\text{T}} \mathbf{b}_j$ and the
	prior 
	$$
	p(\mathbf{B}) \overset{D}{=} \prod_{j=1}^p
	  \mathcal{N}_d(\mathbf{0}_{d \times 1}, \gamma_j \mathbf{I}_d) 
	  \mathbbm{1}\{ \mathbf{b}_j^{\text{T}} \mathbf{b}_j < 1 \}
	$$ 
	is truncated to the unit ball to ensure a non-degenerate correlation matrix.
	
	A mean-field approximation to the posterior of model (\ref{eq:cormodel}) and 
	(\ref{eq:corprior}) is 
	$$
	p(\bm{\Lambda},\mathbf{B},\psi_1, \dots, \psi_p | \mathbf{X}) \approx
	  q(\bm{\Lambda})q(\mathbf{B})q(\psi_1, \dots, \psi_p),
	$$
	such that the Kullback-Leibler divergence of the posterior
  from the (approximate) variational posterior is minimised.
	Here, a slight abuse of notation allows $q$ to refer to 
	different functions depending on the input variable. 
	In general, in mean-field variational Bayes with parameters 
	$\bm{\theta} = \{ \theta_1, \dots, \theta_K \}$, data $\mathbf{X}$, 
  and an assumed factorised approximate posterior: 
  $p(\bm{\theta} | \mathbf{X}) \approx \prod_{k=1}^K q(\theta_k | \mathbf{X})$, 
  results in
  $q(\theta_k | \mathbf{X}) \propto
  \exp \{ \mathbb{E}_{\bm{\theta}_{-k} | \theta_k, \mathbf{X}} 
  \[ \log p( \theta_k | \bm{\theta}_{-k}, \mathbf{X}) \] \}$. 
  For model (\ref{eq:cormodel}) and (\ref{eq:corprior}) this results in:
	\begin{align*}
	  q(\bm{\Lambda}) & \overset{D}{=} \prod_{i=1}^{n} \mathcal{N}_d 
      (\bm{\phi}_i, \bm{\Xi}), \\
	  q(\mathbf{B}) & \overset{D}{=} \prod_{j=1}^{p} \mathcal{N}_d(
	    \bm{\mu}^*_j, \bm{\Omega}^*_j) 
	    \mathbbm{1}\{ \mathbf{b}_j^{\text{T}} \mathbf{b}_j < 1 \}, \\
	  q(\psi_1, \dots, \psi_p) & \overset{D}{=} \prod_{j=1}^p
	    \delta \( \psi_j - \zeta_j \),
	\end{align*}
	with
	\begin{align*}
	  \bm{\phi}_i & = \left\{ \sum_{j=1}^{p} \mathbb{E}(\psi_j)^{-1}
      \[\mathbb{V}(\bar{\mathbf{b}}_j) + \mathbb{E}(\bar{\mathbf{b}}_j) 
      \mathbb{E}(\bar{\mathbf{b}}^{\text{T}}_j)\] + \mathbf{I}_d \right\}^{-1}
      \mathbb{E}(\bar{\mathbf{B}}) \mathbb{E}(\bar{\bm{\Psi}})^{-1} 
      \mathbf{x}_i,\\ 
    \bm{\Xi} & = \left\{ \sum_{j=1}^{p} \mathbb{E}(\psi_j)^{-1}
      \[\mathbb{V}(\bar{\mathbf{b}}_j) + \mathbb{E}(\bar{\mathbf{b}}_j) 
      \mathbb{E}(\bar{\mathbf{b}}_j^{\text{T}})\] + \mathbf{I}_d 
      \right\}^{-1},\\
    \bm{\mu}^*_j &= \[\mathbb{E}(\bm{\Lambda}^{\text{T}}\bm{\Lambda}) +
      \gamma_j^{-1} \mathbf{I}_d\]^{-1}
      \mathbb{E}(\bm{\Lambda}^{\text{T}})\mathbf{x}_j,\\
    \bm{\Omega}^*_j & = \mathbb{E}(\psi_j)
      \[\mathbb{E}(\bm{\Lambda}^{\text{T}}\bm{\Lambda}) +
      \gamma_j^{-1} \mathbf{I}_d\]^{-1}, \\
    \zeta_j & = 1 - \mathbb{E}(\bar{\mathbf{b}}_j^{\text{T}}) 
      \mathbb{E}(\bar{\mathbf{b}}_j) - 
      \text{tr}[\mathbb{V}(\bar{\mathbf{b}}_j)].
	\end{align*}
	The expectations involving $\bm{\Lambda}$ and $\psi_j$ are:
	\begin{align*}
    \mathbb{E}(\bm{\Lambda}^{\text{T}} \bm{\Lambda}) & = n\bm{\Xi} + 
    \bm{\Phi}^{\text{T}} \bm{\Phi},\\
    \mathbb{E}(\bm{\Lambda}) & = \bm{\Phi}, \\
    \mathbb{E}(\psi_j) & = \zeta_j.
  \end{align*}
  
  \subsubsection{Elliptical truncated multivariate Gaussian}
  The expectations and variances involving the $\mathbf{b}_j$ are not available
  in closed form and little bit more involved. They are expectations and
  variances of multivariate elliptical truncated Gaussians.
  We resort to calculation using
  the moment generating function as in \cite{arismendi_multivariate_2017}. This
  results in the following expressions for the expectation and variance:
  \begin{align*}
    \bm{\mu}_j = \mathbb{E}(\bar{\mathbf{b}}_j) & = \bm{\mu}^*_j + L^{-1} 
      \mathbf{v}, \\
    \bm{\Omega}_j = \mathbb{V}(\bar{\mathbf{b}}_j) & = \bm{\Omega}^*_j + L^{-1} 
      \mathbf{V} - L^{-2} \mathbf{v} \mathbf{v}^{\text{T}},
  \end{align*}
  where 
  \begin{align*}
    L & = \sum_{t=0}^{\infty} c_t F_{p + 2t}(r^{-1}), \\
    \mathbf{v} & = \sum_{t=0}^{\infty} \mathbf{c}'_t F_{p + 2t}(r^{-1}), \\
    \mathbf{V} & = \sum_{t=0}^{\infty} \mathbf{C}^{''}_t F_{p + 2t}(r^{-1}), 
  \end{align*}
  are infinite sums that we truncate to approximate the moments. Here, 
  $F_{p}(x)$ denotes the chi-squared distribution function of $x$ with $p$ 
  degrees of freedom.
  
  The coefficients of the sums are calculate recursively:
  \begin{align*}
    c_t & = \frac{1}{2t}\sum_{s=0}^{t - 1} d_{t - s} c_s, \\
    \mathbf{c}'_t & = \frac{1}{2t}\sum_{s=0}^{t - 1} \(\mathbf{d}'_{t - s} c_s +
      d_{t - s} \mathbf{c}'_s \), \\
    \mathbf{C}^{''}_t & = \frac{1}{2t}\sum_{s=0}^{t - 1} 
    \(\mathbf{D}^{''}_{t - s} c_s + \mathbf{d}'_{t - s} 
      (\mathbf{c}'_s)^{\text{T}} +  \mathbf{c}'_s
      (\mathbf{d}'_{t - s})^{\text{T}} + d_{t - s} \mathbf{C}^{''}_s \).
  \end{align*}
  We write $\bm{\Omega}^*_j = \mathbf{Q} \bm{\Lambda} \mathbf{Q}^{\text{T}}$ for 
  the eigen decomposition of $\bm{\Omega}_j^*$ and 
  $z^m_j = r \lambda_j^{-1} (1 - r \lambda_j^{-1})^{m - 1}$. Then, the 
  coefficients $d_{m}$, $\mathbf{d}'_m$, and $\mathbf{D}^{''}_m$ are:
  \begin{align*}
    d_{m} & = \sum_{j=1}^{p + 1} (1 - r \lambda_j^{-1})^m + 
      m (\bm{\mu}_j^{*})^{\text{T}} \mathbf{Q} \bm{\Lambda}^{-1/2}
      \text{diag}(z_j^m) \bm{\Lambda}^{-1/2} \mathbf{Q}^{\text{T}}
      \bm{\mu}_j^{*} , \\
    \mathbf{d}'_m & = -2m \mathbf{Q} \bm{\Lambda}^{1/2}
      \text{diag}(z_j^m) \bm{\Lambda}^{-1/2} \mathbf{Q}^{\text{T}} 
      \bm{\mu}_j^{*},\\
    \mathbf{D}^{''}_m & = 2m \mathbf{Q} \bm{\Lambda}^{1/2}
      \text{diag}(z_j^m) \bm{\Lambda}^{1/2} \mathbf{Q}^{\text{T}}.
  \end{align*}
  Lastly, the first recursive coefficients of the infinite sums are:
  \begin{align*}
    c_0 & = \exp\(-\frac{1}{2} (\bm{\mu}_j^{*})^{\text{T}} 
      (\bm{\Omega}^*_j)^{-1}
      \bm{\mu}_j^* \) \prod_{j=1}^{p+1}r^{1/2}\lambda_j^{-1/2}, \\
    \mathbf{c}'_0 & = - \bm{\mu}^*_j 
      \exp\(-\frac{1}{2} (\bm{\mu}^*_j)^{\text{T}} (\bm{\Omega}^*_j)^{-1}
      \bm{\mu}^*_j \) \prod_{j=1}^{p+1}r^{1/2}\lambda_j^{-1/2}, \\
    \mathbf{C}^{''}_0 & = \(\bm{\mu}^*_j (\bm{\mu}^*_j)^{\text{T}} - 
      \bm{\Omega}^*_j \) 
      \exp\(-\frac{1}{2} (\bm{\mu}^*_j)^{\text{T}} (\bm{\Omega}^*_j)^{-1}
      \bm{\mu}^*_j \) \prod_{j=1}^{p+1}r^{1/2}\lambda_j^{-1/2}.
  \end{align*}
  The remaining $r$ is a free parameter that may be chosen to balance accuracy
  and speed of convergence. \cite{sheil_algorithm_1977} found empirically that 
  $r = 29/32 \min(\lambda_j)$ gives good performance. In a small simulation 
  study we found that truncating the series at $t=50$ gives high enough 
  accuracy for our purposes.
  
  \subsubsection{Variational evidence lower bound}
  The evidence lower bound (ELBO) of model (\ref{eq:cormodel}) and (\ref{eq:corprior})
  is given by:
  \begin{align*}
    \text{ELBO} & = -\frac{n(p + 1)}{2} \log 2\pi + \frac{dn}{2} - 
      \frac{d}{2} \sum_{j=1}^{p + 1} \log \gamma_j -
      \frac{n}{2} \log |\bm{\Delta}| - 
      \frac{1}{2} \sum_{j=1}^{p + 1} \zeta_j^{-1} \sum_{i=1}^n \upsilon_{ij}^2 -
      \frac{n}{2} \sum_{j=1}^{p+1} \zeta_j^{-1} \chi_j \\
      & + \sum_{i=1}^n \bm{\phi}_i^{\text{T}} \mathbf{M} 
      \text{diag}(\zeta_j^{-1}) \bm{\upsilon}_i - 
      \frac{n}{2} \sum_{j=1}^{p+1} \zeta_j^{-1} 
      \text{tr} \( \bm{\Omega}_j \bm{\Xi} \) - 
      \frac{n}{2} \sum_{j=1}^{p+1} \zeta_j^{-1} \bm{\mu}_j^{\text{T}} \bm{\Xi}
      \bm{\mu}_j - \frac{1}{2} \sum_{j=1}^{p+1} \zeta_j^{-1} \sum_{i=1}^n
      \bm{\phi}_i^{\text{T}} \bm{\Omega}_j \bm{\phi}_i \\
      & - \frac{1}{2} \sum_{j=1}^{p+1} \zeta_j^{-1} 
      \sum_{i=1}^n \( \bm{\phi}_i^{\text{T}} \bm{\mu}_j \)^2 - 
      \frac{n}{2} \sum_{j=1}^{p+1} \log \zeta_j - 
      \frac{n}{2} \text{tr} \( \bm{\Delta}^{-1} \bm{\Xi} \) - 
      \frac{1}{2} \sum_{i=1}^n \bm{\phi}_i^{\text{T}} \bm{\Delta}^{-1} 
      \bm{\phi}_i + \frac{n}{2} \log |\bm{\Xi}| \\ 
      & - \sum_{j=1}^{p+1} \log L_p(\mathbf{b}_j) + 
      \sum_{j=1}^{p+1} \log L_q(\mathbf{b}_j) - 
      \frac{d}{2} \sum_{j=1}^{p+1} \log \zeta_j + 
      \frac{1}{2} \sum_{j=1}^{p+1} \log | \bm{\Omega}_j^* | -
      \frac{1}{2} \sum_{j=1}^{p+1} \zeta_j^{-1} \gamma_j^{-1} 
      \text{tr} \( \bm{\Omega}_j \) \\
      & - \frac{1}{2} \sum_{j=1}^{p+1} \zeta_j^{-1} \gamma_j^{-1} 
      \bm{\mu}_j^{\text{T}} \bm{\mu}_j +
      \frac{1}{2} \sum_{j=1}^{p+1}(\bm{\mu}_j^*)^{\text{T}}
      (\bm{\Omega}_j^*)^{-1}\bm{\mu}_j^* - 
      \sum_{j=1}^{p+1} (\bm{\mu}_j^*)^{\text{T}} (\bm{\Omega}_j^*)^{-1}
      \bm{\mu}_j \\
      & + \frac{1}{2} \sum_{j=1}^{p+1} \bm{\mu}_j^{\text{T}} 
      (\bm{\Omega}_j^*)^{-1} \bm{\mu}_j + \frac{1}{2} \sum_{j=1}^{p+1} 
      \text{tr} \( (\bm{\Omega}_j^*)^{-1} \bm{\Omega_j} \).
  \end{align*}
  
  \subsubsection{Ad hoc approach}
  An ad hoc approach to the Bayesian correlation matrix is to freely estimate a 
  general covariance matrix and 
  after estimation apply a correction to ensure that the posterior mean is a
  correlation matrix: $\forall j: \E_{\bar{\mathbf{b}}_j,\bar{\psi}_j | 
  \bar{\mathbf{X}}} \(\mathbf{b}_j^{\text{T}} \mathbf{b}_j + \psi_j\) = 1$.
  To that end we use the VB approximation to write
  $$
  c_j = \E_{\bar{\mathbf{b}}_j,\bar{\psi}_j | 
  \bar{\mathbf{X}}} \(\mathbf{b}_j^{\text{T}} \mathbf{b}_j + \psi_j\) \approx
  \bm{\mu}_j^{\text{T}}\bm{\mu}_j + \text{tr}\( \bm{\Omega}_j \) +
  \frac{\zeta_j}{n/2 + \kappa_j - 1},
  $$
  and use the rescaled variational posterior for prediction:
  \begin{align*}
    q(\bar{\mathbf{B}}) & \overset{D}{=} \prod_{j=1}^{p + 1} \mathcal{N}_d 
      (c_j^{-1/2}\bm{\mu}_j, c_j^{-1}\bm{\Omega}_j) \\
    q(\bar{\bm{\Psi}}) & \overset{D}{=} \prod_{j=1}^{p+1} \Gamma^{-1}
      \(\frac{n + m + d}{2} + \kappa_j, c_j^{-1}\zeta_j\).
  \end{align*}
  Inspection of the prediction rule then gives:
  \begin{align*}
    \mathbb{E}(\tilde{y}|\tilde{\mathbf{x}}) & = \tilde{\mathbf{x}}^{\text{T}}
    \diag (c_j)\bm{\Psi}^{-1} \diag (c_j^{-1/2}) \mathbf{B}^{\text{T}}
    \left\{ \mathbf{B} \diag (c_j^{-1/2})\diag (c_j) \bm{\Psi}^{-1} 
    \diag (c_j^{-1/2}) \mathbf{B}^{\text{T}} + \mathbf{I}_d\right\}^{-1}
    \bm{\beta}c_{p+1}^{-1/2} \\
    & = \tilde{\mathbf{x}}^{\text{T}} \diag \[ \(c_j/c_{p+1}\)^{1/2} \]
    \tilde{\bm{\beta}}, \,\, j=1,\dots, p,
  \end{align*}
  which shows that the ad hoc approach is a rescaling of the original 
  prediction rule, with scaling factor the ratio of
  feature to outcome posterior standard deviation.
  
  \section{Logistic regression}\label{sec:logistic}
	In the case of sums of $N_i$ disjoint binary events $y_i$, we consider the
	logistic model for the outcomes. In logistic models, we cannot center
	the data to remove any fixed mean effects from the model. We therefore include
	a mean/intercept term $\bar{\bm{\beta}} = \begin{bmatrix} \beta_0 & 
	\bm{\beta}^{\text{T}} \end{bmatrix}^{\text{T}}$ in the observational model and
	introduce the following Bayesian
  factor regression model for binomial outcomes $y_i$:
	\begin{align*}
	  y | \bm{\lambda}, \bar{\bm{\beta}} & \sim \mathcal{B}\left(N,
	    \expit(\beta_0 + \bm{\beta}^{\text{T}}\bm{\lambda})\right), \\
	  \eta | \bm{\lambda}, \bar{\bm{\beta}} & \sim \mathcal{PG}(N,
	    \beta_0 + \bm{\beta}^{\text{T}}\bm{\lambda}), \\
	  \mathbf{x} | \bm{\lambda}, \mathbf{B}, \psi_1, \dots, \psi_p & \sim 
	    \mathcal{N}_p(\mathbf{B}^{\text{T}} \bm{\lambda}, \bm{\Psi}), \\
	  \bm{\lambda} & \sim \mathcal{N}_d (\mathbf{0}_{d}, \mathbf{I}_d), \\
	  \bm{\beta} & \sim \mathcal{N}_d (\mathbf{0}_{d}, \gamma_{\bar{p}} 
	    \mathbf{I}_d), \\
	  \beta_0 & \sim 1, \\
	  \mathbf{B} | \psi_1, \dots, \psi_{p} & \sim 
  	  \prod_{j=1}^{p} \mathcal{N}_{d}(\mathbf{0}_{d}, 
  	  \psi_{j} \gamma_{j} \mathbf{I}_d), \\
  	\psi_1, \dots, \psi_{p} & \sim \prod_{j=1}^{p} 
  	  \Gamma^{-1}(\kappa_j, \nu_j),
  \end{align*}	
	where we have introduced additional latent variables
	$\eta$, with
	$\mathcal{PG} (N, \delta)$, $N > 0$, $\delta \in \mathbb{R}$, the
  P{\'o}lya-Gamma distribution \cite[]{polson_bayesian_2013}. Note that the
  intercept $\beta_0$ is given a flat prior and is therefore not directly 
  shrunken.
  
  Similar to the linear case, we switch to the joint notation with
  $\bar{\mathbf{X}} = \begin{bmatrix} \mathbf{X} & \mathbf{y} - \mathbf{N}/2
  \end{bmatrix}$, $\mathbf{N} = \begin{bmatrix} N_1 & \cdots & N_n 
  \end{bmatrix}^{\text{T}}$,
  $\bar{\mathbf{B}} = \begin{bmatrix} \mathbf{B} & \bm{\beta} \end{bmatrix}$,
  $\bar{\mathbf{H}}=\begin{bmatrix} \mathbf{1}_{n \times p} &
  \bm{\eta}\end{bmatrix}$, and
  $$
	\bar{\bm{\Psi}}=\begin{bmatrix} 
	\bm{\Psi} & \mathbf{0}_{p \times 1} \\
	\mathbf{0}_{1 \times p} & 1
	\end{bmatrix}.
	$$ 
	In addition, we introduce a slight abuse of notation by letting
  $\bar{\bm{\eta}}_i$ and $\bar{\bm{\eta}}_j$ denote the $i$th row and $j$th
  column of $\bar{\mathbf{H}}$, respectively. Extension to unlabeled features
  is straightforward by considering additional unobserved outcomes 
  $z_i$, $i=n+1, \dots, n + m$ that follow the same model as the observed 
  outcomes.
  
  \subsection{Gibbs sampler}
  The full conditionals for $\bm{\eta}$, $\bm{\Lambda}$, $\bar{\bm{\beta}}$, 
  $\mathbf{B}$, and
  $\psi_1, \dots, \psi_p$ are derived in a similar way as in the linear model.
  The full conditional for $\bm{\eta}$ is the same as in the prior:
	\begin{align*}
	p(\bm{\eta} | \mathbf{y}, \bm{\Lambda}, \bar{\bm{\beta}}) & =
	  \prod_{i=1}^{n+m} p(\eta_i | \bm{\lambda}_i, \bar{\bm{\beta}}),
	\end{align*}
	so that we have
	$$
	\bm{\eta} | \mathbf{y}, \bm{\Lambda}, \bar{\bm{\beta}} \sim 
	\prod_{i=1}^{n+m} \mathcal{PG} 
	(N_i, \beta_0 + \bm{\beta}^{\text{T}} \bm{\lambda}_i)
	$$
  
  For $\bm{\Lambda}$ we have:
  \begin{align*}
	  p(\bm{\Lambda} | \mathbf{X}, \mathbf{y}, \mathbf{z}, \bm{\eta}, 
	    \bar{\mathbf{B}}, 
  	  \psi_1, \dots \psi_p) & \propto \prod_{i=1}^{n + m} 
  	  p(y_i | \bm{\lambda}_i, \bar{\bm{\beta}})
  	  p(\eta_i | \bm{\lambda}_i, \bar{\bm{\beta}})
  	  p(\mathbf{x}_i | \bm{\lambda}_i, \mathbf{B}, \psi_1, 
  	  \dots \psi_p) p(\bm{\lambda}_i) \\
    & \propto \prod_{i=1}^{n + m} 
      \frac{\exp(\beta_0 + \bm{\beta}^{\text{T}} \bm{\lambda}_i)^{y_i}}
      {\[\exp(\beta_0 + \bm{\beta}^{\text{T}} \bm{\lambda}_i) + 1\]^{N_i}} \\
    & \quad \quad \quad \exp\[-\eta_i(\beta_0 + \bm{\beta}^{\text{T}} 
      \bm{\lambda}_i)^2/2\] \cosh \[(\beta_0 + 
      \bm{\beta}^{\text{T}} \bm{\lambda}_i)/2 \]^{N_i}\\
    & \quad \quad \quad \exp \[ -\frac{1}{2} (\mathbf{x}_i - 
      \mathbf{B}^{\text{T}} 
      \bm{\lambda}_i)^{\text{T}} 
      \bm{\Psi}^{-1} (\mathbf{x}_i - \mathbf{B}^{\text{T}} 
      \bm{\lambda}_i)
      - \frac{1}{2}\bm{\lambda}_i^{\text{T}} \bm{\lambda}_i \] \\
    & \propto \prod_{i=1}^{n + m} 
      \exp \bigg\{ -\frac{1}{2} \bm{\lambda}_i^{\text{T}}
      (\mathbf{B} \bm{\Psi}^{-1} \mathbf{B}^{\text{T}} + 
      \eta_i \bm{\beta} \bm{\beta}^{\text{T}} + \mathbf{I}_d)
      \bm{\lambda}_i \\
    & \quad \quad \quad + \bm{\lambda}_i^{\text{T}}
      \[\mathbf{B} \bm{\Psi}^{-1} \mathbf{x}_i + (y_i - N_i/2 - \eta_i 
      \beta_0)\bm{\beta}\]\bigg\},
	\end{align*}
	so that 
	\begin{align*}
	  \bm{\Lambda} | \mathbf{X}, \mathbf{y}, \mathbf{z}, \bm{\eta}, 
	    \bar{\mathbf{B}}, \psi_1, \dots \psi_p & \sim \mathcal{N}_d 
	    \big(\mathbf{B} \bm{\Psi}^{-1} \mathbf{B}^{\text{T}} + 
      \eta_i \bm{\beta} \bm{\beta}^{\text{T}} + \mathbf{I}_d)^{-1}
      \[\mathbf{B} \bm{\Psi}^{-1} \mathbf{x}_i + (y_i - N_i/2 - \eta_i 
      \beta_0)\bm{\beta}\], \\
    & \quad \quad \quad (\mathbf{B} \bm{\Psi}^{-1} \mathbf{B}^{\text{T}} + 
      \eta_i \bm{\beta} \bm{\beta}^{\text{T}} + \mathbf{I}_d)^{-1}\big)
	\end{align*}
	
	For $\bar{\bm{\beta}}$ we have:
  \begin{align*}
	  p(\bar{\bm{\beta}} | \mathbf{y}, \bm{\eta}, \bm{\Lambda}) & \propto
	    \prod_{i=1}^{n + m} p(y_i | \bm{\lambda}_i, \bar{\bm{\beta}})
  	  p(\eta_i | \bm{\lambda}_i, \bar{\bm{\beta}})
  	  p(\bm{\beta})p(\beta_0) \\
    & \propto \prod_{i=1}^{n + m}
      \frac{\exp(\beta_0 + \bm{\beta}^{\text{T}} \bm{\lambda}_i)^{y_i}}
      {\[\exp(\beta_0 + \bm{\beta}^{\text{T}} \bm{\lambda}_i) + 1\]^{N_i}} \\
    & \quad \quad \quad \exp\[-\eta_i(\beta_0 + \bm{\beta}^{\text{T}} 
      \bm{\lambda}_i)^2/2\] \cosh \[(\beta_0 + 
      \bm{\beta}^{\text{T}} \bm{\lambda}_i)/2 \]^{N_i} \\
    & \quad \quad \quad \exp \( \frac{\gamma_{\bar{p}}^{-1}}{2} 
      \bm{\beta}^{\text{T}} \bm{\beta}\) \\
    & \propto \exp \bigg\{ - \frac{1}{2} \bar{\bm{\beta}}^{\text{T}}
      \begin{bmatrix} \sum_{i=1}^{n+m} \eta_i & \bm{\eta}^{\text{T}} 
      \bm{\Lambda} \\
      \bm{\Lambda}^{\text{T}} \bm{\eta} & \bm{\Lambda}^{\text{T}}
      \text{diag} (\eta_i) \bm{\Lambda} + \gamma_{\bar{p}}^{-1} \mathbf{I}_d
      \end{bmatrix} \bar{\bm{\beta}} \\
    & \quad \quad \quad + \bar{\bm{\beta}}^{\text{T}} 
      \begin{bmatrix} \sum_{i=1}^{n+m} (y_i - N_i/2) \\
      \bm{\Lambda}^{\text{T}} (\mathbf{y} - \mathbf{N}/2)
      \end{bmatrix}\bigg\}. 
	\end{align*}
	This gives:
	\begin{align*}
	\bar{\bm{\beta}} | \mathbf{y}, \bm{\eta}, \bm{\Lambda} & \sim \mathcal{N}_{d+1}
	  \bigg( \begin{bmatrix} \sum_{i=1}^{n+m} \eta_i & \bm{\eta}^{\text{T}} 
    \bm{\Lambda} \\
    \bm{\Lambda}^{\text{T}} \bm{\eta} & \bm{\Lambda}^{\text{T}}
    \text{diag} (\eta_i) \bm{\Lambda} + \gamma_{\bar{p}}^{-1} \mathbf{I}_d
    \end{bmatrix}^{-1}\begin{bmatrix} \sum_{i=1}^{n+m} (y_i - N_i/2) \\
    \bm{\Lambda}^{\text{T}} (\mathbf{y} - \mathbf{N}/2)
    \end{bmatrix}, \\
  & \quad \quad \quad \begin{bmatrix} \sum_{i=1}^{n+m} \eta_i & 
    \bm{\eta}^{\text{T}} 
    \bm{\Lambda} \\
    \bm{\Lambda}^{\text{T}} \bm{\eta} & \bm{\Lambda}^{\text{T}}
    \text{diag} (\eta_i) \bm{\Lambda} + \gamma_{\bar{p}}^{-1} \mathbf{I}_d
    \end{bmatrix}^{-1}\bigg)
	\end{align*}

  Next, we derive the conditional for $\mathbf{B}$:
  \begin{align*}
	  p(\mathbf{B} | \mathbf{X}, \bm{\Lambda}, \psi_1, \dots \psi_p) & \propto 
	    \prod_{i=1}^{n + m} 
  	  p(\mathbf{x}_i | \bm{\lambda}_i, \mathbf{B}, \psi_1, 
  	  \dots \psi_p) \prod_{j=1}^{p} p(\mathbf{b}_j | \psi_j) \\
  	& = \prod_{j=1}^p p(\mathbf{x}_j | \bm{\Lambda}, \mathbf{b}_j, \psi_j) 
  	  p(\mathbf{b}_j | \psi_j) \\
  	& \propto \prod_{j=1}^p 
  	  \exp \[ \frac{\psi_j^{-1}}{2} 
  	  (\mathbf{x}_j - \bm{\Lambda} \mathbf{b}_j)^{\text{T}} 
  	  (\mathbf{x}_j - \bm{\Lambda} \mathbf{b}_j) - 
  	  \frac{\gamma_{j}^{-1} \psi_{j}^{-1}}{2}
  	  \mathbf{b}_j^{\text{T}} \mathbf{b}_j \] \\
    & \propto \prod_{j=1}^p \exp \bigg\{ -\frac{1}{2} 
      \[ \mathbf{b}_{j} - ( \bm{\Lambda}^{\text{T}}\bm{\Lambda} + 
  	  \gamma^{-1}_{j}\mathbf{I}_d)^{-1}
  	  \bm{\Lambda}^{\text{T}} \mathbf{x}_{j}\]^{\text{T}} \\
    & \quad \quad \quad 
      \psi_{j}^{-1} (\bm{\Lambda}^{\text{T}}\bm{\Lambda} + 
  	  \gamma^{-1}_{j}\mathbf{I}_d) \[ \mathbf{b}_{j} - 
  	  ( \bm{\Lambda}^{\text{T}}\bm{\Lambda} + 
  	  \gamma^{-1}_{j}\mathbf{I}_d)^{-1}
  	  \bm{\Lambda}^{\text{T}} \mathbf{x}_{j}\] \bigg\},
	\end{align*}
	which gives
	\begin{align*}
  	\mathbf{B} | \mathbf{X}, \bm{\Lambda}, \psi_1, \dots \psi_p
  	  \sim \prod_{j=1}^{p}
    	\mathcal{N}_{d} \( ( \bm{\Lambda}^{\text{T}}\bm{\Lambda} + 
    	\gamma^{-1}_{j}\mathbf{I}_d)^{-1}
    	\bm{\Lambda}^{\text{T}} \mathbf{x}_{j},
    	\psi_{j} (\bm{\Lambda}^{\text{T}}\bm{\Lambda} + 
    	  \gamma^{-1}_{j}\mathbf{I}_d)^{-1}\). 
	\end{align*}
  
  For the $\psi_{j}$, we derive:
	\begin{align*}
	  p(\psi_1, \dots \psi_{p} | \mathbf{x}, \bm{\Lambda}, 
  	  \mathbf{b}) & \propto \prod_{i=1}^{n + m} 
  	  p(\mathbf{x}_i | \bm{\lambda}_i, \mathbf{b}, \psi_1, 
  	  \dots \psi_{p}) \prod_{j=1}^{p} 
  	  p(\mathbf{b}_{j} | \psi_{j})
  	  p(\psi_{j}) \\
  	& = \prod_{j=1}^{p} p(\mathbf{x}_{j} | 
  	  \bm{\Lambda}, \mathbf{b}_{j}, \psi_{j})
  	  p(\mathbf{b}_{j} | \psi_{j})
  	  p(\psi_{j}) \\
  	& \propto \prod_{j=1}^{p} 
  	  \psi_{j}^{-(\frac{n + m + d}{2} + \kappa_{j}) - 1} \\
  	& \quad \quad \quad \exp \left\{ -\psi_{j}^{-1} \[\frac{1}{2} 
  	   (\mathbf{x}_{j} - 
  	  \bm{\Lambda}\mathbf{b}_{j})^{\text{T}}
  	  (\mathbf{x}_{j} - 
  	  \bm{\Lambda}\mathbf{b}_{j}) + \frac{\gamma^{-1}_{j}}{2}
  	  \mathbf{b}_{j}^{\text{T}}\mathbf{b}_{j} + 
  	  \nu_{j}\] \right\},
	\end{align*}
	to arrive at
	\begin{align*}
	\psi_1, \dots \psi_{p} | \mathbf{x}, \bm{\Lambda}, 
	  \mathbf{b} \sim \prod_{j=1}^{p} 
	  \Gamma^{-1}\(\frac{n + m + d}{2} + \kappa_{j},
	  \frac{1}{2}(\mathbf{x}_{j} - 
	  \bm{\Lambda}\mathbf{b}_{j})^{\text{T}}
	  (\mathbf{x}_{j} - 
	  \bm{\Lambda}\mathbf{b}_{j}) + 
	  \frac{\gamma_{j}^{-1}}{2}\mathbf{b}_{j}^{\text{T}}
	  \mathbf{b}_{j} + \nu_j \). 
	\end{align*}
  A Gibbs sample from the posterior is now generated by sequentially sampling 
  from these full conditionals.
	
	\subsection{Variational inference}
	If, as before, we consider the posterior factorisation:
	$$
	q(\bm{\eta})q(\bm{\Lambda})q(\bar{\bm{\beta}},\mathbf{B})q(\bm{\psi})
	q(\mathbf{z})
	$$
	the corresponding variational posterior is:
  \begin{subequations}\label{eq:logvariationaldistributions}
    \begin{align*}
      q(\eta_1, \dots, \eta_{n+m}) & \overset{D}{=} \prod_{i=1}^{n+m}
        \mathcal{PG}(N_i, \delta_i), \\
      q(\bm{\Lambda}) & \overset{D}{=} \prod_{i=1}^{n + m} \mathcal{N}_d 
        (\bm{\phi}_i, \bm{\Xi}_i), \\
      q(\bar{\bm{\beta}}) & \overset{D}{=} \mathcal{N}_{d+1} 
        (\bm{\mu}_{\bar{p}}, \bm{\Omega}_{\bar{p}}), \\
      q(\mathbf{B}) & \overset{D}{=} \prod_{j=1}^{p} \mathcal{N}_d 
        (\bm{\mu}_j, \bm{\Omega}_j), \\
      q(\psi_1, \dots, \psi_{p}) & \overset{D}{=} 
        \prod_{j=1}^{p} \Gamma^{-1}
        \(\frac{n + m + d}{2} + \kappa_j, \zeta_j\), \\
      q(\mathbf{z}) & \overset{D}{=} \prod_{i=n + 1}^{n + m} \mathcal{B}
        \left(N_i, \upsilon_i)\right),
    \end{align*}
  \end{subequations}
  with parameters:
  \begin{subequations}
    \begin{align}
      \delta_i & = \Big\{ \mathbb{E}(\bm{\beta}^{\text{T}}) 
        \mathbb{V}(\bm{\lambda}_i) \mathbb{E}(\bm{\beta}) + 
        \left[ \mathbb{E}(\bm{\lambda}_i^{\text{T}}) 
        \mathbb{E}(\bm{\beta})\right]^2 + 
        \text{tr} \left[ \mathbb{V}(\bm{\beta})
        \mathbb{V}(\bm{\lambda}_i)\right] + 
        \mathbb{E}(\bm{\lambda}_i^{\text{T}}) 
        \mathbb{V}(\bm{\beta}) \mathbb{E}(\bm{\lambda}_i) \\
      & \,\,\,\,\,\, \,\,\, 2 \mathbb{E}(\bm{\lambda}_i^{\text{T}}) 
        \mathbb{C}\text{ov}(\bm{\beta},\beta_0) + 
        2 \mathbb{E}(\bm{\lambda}_i^{\text{T}}) \mathbb{E}(\bm{\beta})
        \mathbb{E}(\beta_0) + \mathbb{E}(\beta_0)^2 + \mathbb{V}(\beta_0) 
        \Big\}^{1/2}, \\
      \bm{\phi}_i & = \left\{ \sum_{\bar{j}=1}^{\bar{p}} 
        \mathbb{E}(\bar{\eta}_{i{\bar{j}}})
        \mathbb{E}(\bar{\psi}_{{\bar{j}}}^{-1}) 
        \[\mathbb{V}(\bar{\mathbf{b}}_{\bar{j}}) + 
        \mathbb{E}(\bar{\mathbf{b}}_{\bar{j}}) 
        \mathbb{E}(\bar{\mathbf{b}}^{\text{T}}_{\bar{j}})\] + 
        \mathbf{I}_d \right\}^{-1}
        \big[\mathbb{E}(\bar{\mathbf{B}}) \mathbb{E}(\bar{\bm{\Psi}}^{-1}) 
        \tilde{\mathbf{x}}_i - \mathbb{E}(\eta_i)\mathbb{E}(\beta_0)
        \mathbb{E}(\bm{\beta}) \\
      & \,\,\,\,\,\, \,\,\, - \mathbb{E}(\eta_i) \mathbb{C}\text{ov}
        (\bm{\beta},\beta_0)\big],\\ 
      \bm{\Xi}_i & = \left\{ \sum_{\bar{j}=1}^{\bar{p}} 
        \mathbb{E}(\bar{\eta}_{i{\bar{j}}})
        \mathbb{E}(\bar{\psi}_{{\bar{j}}}^{-1}) 
        \[\mathbb{V}(\bar{\mathbf{b}}_{\bar{j}}) + 
        \mathbb{E}(\bar{\mathbf{b}}_{\bar{j}}) 
        \mathbb{E}(\bar{\mathbf{b}}^{\text{T}}_{\bar{j}})\] + 
        \mathbf{I}_d \right\}^{-1},\\
      \bm{\mu}_j &= \[\mathbb{E}(\bm{\Lambda}^{\text{T}})
        \mathbb{E}(\bm{\Lambda}) + \sum_{i=1}^{n+m}\mathbb{V}(\bm{\lambda}_i) +
        \gamma_{j}^{-1} \mathbf{I}_d\]^{-1}
        \mathbb{E}(\bm{\Lambda}^{\text{T}})\tilde{\mathbf{x}}_j, \\
      \bm{\Omega}_j & = \mathbb{E}(\psi_{j}^{-1})^{-1}
        \[\mathbb{E}(\bm{\Lambda}^{\text{T}})
        \mathbb{E}(\bm{\Lambda}) + \sum_{i=1}^{n+m}\mathbb{V}(\bm{\lambda}_i) +
        \gamma_{j}^{-1} \mathbf{I}_d\]^{-1},\\
      \bm{\mu}_{\bar{p}} &= \begin{bmatrix}
        \sum_{i=1}^{n+m} \mathbb{E}(\eta_i) & \mathbb{E}(\bm{\eta})^{\text{T}}
        \mathbb{E}(\bm{\Lambda}) \\
      \mathbb{E}(\bm{\Lambda})^{\text{T}}\mathbb{E}(\bm{\eta}) &
        \sum_{i=1}^{n+m} \mathbb{E}(\eta_{i}) 
        \left[ \mathbb{V}(\bm{\lambda}_i) + \mathbb{E}(\bm{\lambda}_i)
        \mathbb{E}(\bm{\lambda}^{\text{T}}_i) \right] + 
        \gamma_{\bar{p}}^{-1} \mathbf{I}_d
        \end{bmatrix}^{-1} \\
      & \,\,\,\,\,\, \,\,\,
        \begin{bmatrix} \mathbf{1}_{1 \times (n + m)} \\
        \mathbb{E}(\bm{\Lambda}^{\text{T}})
        \end{bmatrix} \tilde{\mathbf{x}}_{{\bar{p}}},\\
      \bm{\Omega}_{\bar{p}} & = 
        \begin{bmatrix}
        \sum_{i=1}^{n+m} \mathbb{E}(\eta_i) & \mathbb{E}(\bm{\eta})^{\text{T}}
        \mathbb{E}(\bm{\Lambda}) \\
        \mathbb{E}(\bm{\Lambda})^{\text{T}}\mathbb{E}(\bm{\eta}) &
        \sum_{i=1}^{n+m} \mathbb{E}(\eta_{i}) 
        \left[ \mathbb{V}(\bm{\lambda}_i) + \mathbb{E}(\bm{\lambda}_i)
        \mathbb{E}(\bm{\lambda}^{\text{T}}_i) \right] + 
        \gamma_{\bar{p}}^{-1} \mathbf{I}_d
        \end{bmatrix}^{-1}, \\
      \zeta_j & = \mathbf{x}_j^{\text{T}} \mathbf{x}_j/2 - 
        \mathbb{E}(\bar{\mathbf{b}}_j^{\text{T}})\mathbb{E}
        (\bm{\Lambda}^{\text{T}})
        \mathbf{x}_j + \text{tr}\[\mathbb{E}(\bm{\Lambda}^{\text{T}})
        \mathbb{E}(\bm{\Lambda})\mathbb{V}(\bar{\mathbf{b}}_j)\]/2 \\
      & \,\,\,\,\,\, \,\,\, +
        \text{tr}\[\sum_{i=1}^{n+m}\mathbb{V}(\bm{\lambda}_i)
        \mathbb{V}(\bar{\mathbf{b}}_j)\]/2 + 
        \mathbb{E}(\bar{\mathbf{b}}_j^{\text{T}})
        \mathbb{E}(\bm{\Lambda}^{\text{T}}) \mathbb{E}(\bm{\Lambda})
        \mathbb{E}(\bar{\mathbf{b}}_j)/2 \\
      & \,\,\,\,\,\, \,\,\, + \mathbb{E}(\bar{\mathbf{b}}_j^{\text{T}})
        \sum_{i=1}^{n+m}\mathbb{V}(\bm{\lambda}_i)
        \mathbb{E}(\bar{\mathbf{b}}_j)/2 + \gamma_j^{-1}\mathbb{E}
        (\mathbf{b}_j^{\text{T}})\mathbb{E}(\mathbf{b}_j)/2 +
        \gamma_j^{-1}\text{tr}\[\mathbb{V}(\mathbf{b}_j)\]/2 + \nu_j, \\
      \upsilon_{i} & = \text{expit}\left[
        \mathbb{E}(\bm{\beta}^{\text{T}})
        \mathbb{E}(\bm{\lambda}_i) + \mathbb{E}(\beta_0)\right],
    \end{align}
  \end{subequations}
  where
  $$
  \tilde{\mathbf{X}} = \begin{bmatrix}
    \mathbf{X} & 
    \begin{bmatrix}
      \begin{bmatrix} 
        \mathbf{y} \\
        \mathbb{E}(\mathbf{z})
      \end{bmatrix} - \mathbf{N}/2
    \end{bmatrix}
  \end{bmatrix}
  $$
  and the expectations and variances are as follows:
  \begin{align*}
    \mathbb{E}(\bar{\psi}_{j}^{-1}) & = \(\frac{n + m + d}{2} + \kappa_j\)/
      \zeta_j,\\
    \mathbb{E}(\bar{\psi}_{\bar{p}}^{-1}) & = 1,\\
    \mathbb{V}(\bar{\mathbf{b}}_{\bar{j}}) & = \bm{\Omega}_{\bar{j}},\\
    \mathbb{E}(\bar{\mathbf{b}}_{\bar{j}}) & = \bm{\mu}_{\bar{j}}, \\
    \mathbb{E}(\bm{\Lambda}) & = \bm{\Phi},\\
    \mathbb{V}(\bm{\lambda}_i)& = \bm{\Xi}_i, \\
    \mathbb{E}(z_{i}) & = N_i \upsilon_{i},\\
    \mathbb{E}(\eta_i) & = N_i \tanh(\delta_i/2)/(2\delta_i),
  \end{align*}
  with $\bm{\Phi} = \begin{bmatrix} \bm{\phi}_1 & \cdots \bm{\phi}_n 
  \end{bmatrix}^{\text{T}}$.
  
  \subsection{Posterior expectation}
  The posterior expectation $\mathbb{E}(\tilde{y} | \tilde{\mathbf{x}})$ 
  for new data $\tilde{y},\tilde{\mathbf{x}}$ is not available in closed form. 
  For the case $N=1$ It is given by 
  $$
  \mathbb{E} \left\{ \text{expit} \left[ \bm{\beta}^{\text{T}} 
  \left(\mathbf{B} \bm{\Psi}^{-1} \mathbf{B}^{\text{T}} + 
  \eta \bm{\beta} \bm{\beta}^{\text{T}} + \mathbf{I}_d \right)^{-1}
  \mathbf{B}\bm{\Psi}^{-1} \tilde{\mathbf{x}}\right] \right\},
  $$
  where the expectation is with respect to 
  $p(\mathbf{B}, \bm{\beta}, \bm{\Psi}, \eta | \mathbf{x})$.
	% In the case of logistic regression, the conditional expectation of the
	% observational model $\mathbb{E}(y | \mathbf{x}) =
	% \mathbb{E} \[m \cdot \expit(\bm{\beta}^{\text{T}}\bm{\lambda} ) | x\]$ is not
	% available in closed form. 
	% 
	An approximation to second order is:
	$$
	\mathbb{E}(\tilde{y} | \tilde{\mathbf{x}}) \approx \left\{ \expit(c) +
	s \cdot \expit(c)\[1 - \expit(c)\]^2/2 -
	s \cdot \expit(c)^2\[1 - \expit(c)\]/2 \right\},
	$$
	with 
	$$
	c=\mathbb{E} \left[\bm{\beta}^{\text{T}} 
  \left(\mathbf{B} \bm{\Psi}^{-1} \mathbf{B}^{\text{T}} + 
  \eta \bm{\beta} \bm{\beta}^{\text{T}} + \mathbf{I}_d \right)^{-1}
  \mathbf{B}\bm{\Psi}^{-1} \tilde{\mathbf{x}}\right]
  $$ 
  and 
  $$
  s=\mathbb{V}
  \left[ \bm{\beta}^{\text{T}} 
  \left(\mathbf{B} \bm{\Psi}^{-1} \mathbf{B}^{\text{T}} + 
  \eta \bm{\beta} \bm{\beta}^{\text{T}} + \mathbf{I}_d \right)^{-1}
  \mathbf{B}\bm{\Psi}^{-1} \tilde{\mathbf{x}}\right].
  $$
  Just as in the linear
  case, $\mathbb{E}(\tilde{y} | \tilde{\mathbf{x}})$ depends not just on
	$\bm{\beta}$, but also on $\mathbf{B}$ and $\bm{\Psi}$, such that additional
	observations on $\mathbf{x}$ might benefit prediction. $c$ and $s$ may be 
	approximated in a similar way to the linear model. That is, we either
	use a Taylor approximation or use Monte Carlo samples from the posterior
	to approximate.
	
	An additional difficulty compared to the linear case is the additional
	expectation of $\eta$, with density:
	$$
	p(\eta | 1, 0)(1 + \eta \bm{\beta}^{\text{T}}\bm{\beta})^{-1/2}
	\exp \left[ \frac{1}{8} \frac{\eta (\bm{\beta}^{\text{T}}\bm{\beta})^2}
	{1 + \eta \bm{\beta}^{\text{T}}\bm{\beta}}\right]
	\exp \left( \frac{\bm{\beta}^{\text{T}}\bm{\beta}}{8} \right),
	$$
	where $p(\eta | 1, 0)$ is the density of a $\mathcal{PG}(1,0)$ distributed
	variable. As far as we are aware, this distribution does not allow for a 
	closed-form expectation, but the distribution is very cheap to sample from, 
	since it only requires sampling from the inner product of a Gaussian 
	$\bm{\beta}^{\text{T}}\bm{\beta}$, which is a generalised chi-square variable,
	and sampling from the $\mathcal{PG}(1,0)$, for which efficient sampling 
	schemes exist.
	
	\subsection{Evidence lower bound}
	Here we give the variational evidence lower bound for the logistic model.
  Let $\tau_j = [(n + m + d)/2 + \kappa_j]/(2\zeta_j)$, then:
  \begin{align*}
    \text{ELBO}= & -\frac{np}{2} \log 2 \pi + 
      \frac{(n + m)d + p(n + m) + dp + d + 1}{2} + 
      \sum_{i=1}^n \log \binom{N_i}{y_i} \\
    & + \sum_{j=1}^{p} \bigg[ \log \Gamma \( \frac{(n + m + d)}{2} + \kappa_j \) -
      \log \Gamma(\kappa_j) - \frac{d}{2} \psi \(\frac{(n + m + d)}{2} + 
      \kappa_j\) \\
    & \quad \quad \quad + \frac{d}{2}\log \gamma_j + \kappa_j \(1 + \log 
      \nu_j \) \bigg] \\
    & - \sum_{j=1}^{p} \left[\left( \frac{n + m - d}{2} + \kappa_j \right) 
      \log \zeta_j 
      + \frac{1}{2} \nu_j \tau_j \right] \\
    & + \sum_{j=1}^p \[\frac{1}{2} \log |\bm{\Omega}_j| -
      \tau_j \gamma_j^{-1} \text{tr}\(\bm{\Omega}_j\) - 
      \sum_{i=1}^{n + m} \tau_j \text{tr}\(\bm{\Xi}_i\bm{\Omega}_j\) - 
      \tau_j \text{tr}\(\bm{\Phi}^{\text{T}}\bm{\Phi}\bm{\Omega}_j\)\] \\
    & + 2 \text{tr} \[\text{diag} \( \tau_j \) 
      \tilde{\mathbf{X}}^{\text{T}} \bm{\Phi} \mathbf{M}\]
      - \sum_{i=1}^{n + m}\text{tr} \[\text{diag} \( \tau_j \) 
      \mathbf{M}^{\text{T}}\bm{\Xi}_i\mathbf{M}\] -
      \text{tr} \[\text{diag} \( \tau_j \gamma_j^{-1} \) 
      \mathbf{M}^{\text{T}}\mathbf{M}\] \\
    & - \text{tr} \[\text{diag} \( \tau_j \) 
      \mathbf{M}^{\text{T}}\bm{\Phi}^{\text{T}}\bm{\Phi}\mathbf{M}\] -
      \text{tr} \[\text{diag} \( \tau_j \) 
      \tilde{\mathbf{X}}^{\text{T}}\tilde{\mathbf{X}}\] \\  
    & + \frac{1}{2}\sum_{i=1}^{n+m} \log |\bm{\Xi}_i| - 
      \frac{1}{2} \sum_{i=1}^{n+m} \text{tr}( \bm{\Xi}_i) -
      \frac{1}{2}\text{tr} \(\bm{\Phi}^{\text{T}} \bm{\Phi}\) \\
    & - \frac{\gamma_{p+1}^{-1}}{2} \text{tr} 
      \[ (\bm{\Omega}_{p+1})_{-1,-1} \] -
      \frac{\gamma_{p+1}^{-1}}{2} (\bm{\mu}_{p+1}^{\text{T}})_{-1}
      (\bm{\mu}_{p+1})_{-1} + \frac{1}{2}\log |\bm{\Omega}_{p+1}| \\
    & - \sum_{i=n+1}^{n+m} N_i \left[ (1 - \upsilon_i)\log (1 - \upsilon_i)
      + \upsilon_i \log \upsilon_i \right] \\
    & + \sum_{i=1}^{n+m} N_i \left\{ (\upsilon_i - \frac{1}{2}) 
      \left[ \begin{bmatrix} 1 & \bm{\phi}_i^{\text{T}} \end{bmatrix}
      \bm{\mu}_{p+1} - \delta_i \right] - 
      \log \left[ 1 + \exp ( \delta_i ) \right]\right\} \\
    & + \sum_{i=1}^{n+m} 
      \frac{ N_i \text{tanh} (\delta_i/2)}{4 \delta_i} 
      \bigg[ \delta_i^2 - \bm{\mu}_{p+1}^{\text{T}} 
      \begin{bmatrix} 
      1 & \bm{\phi}_i^{\text{T}} \\
      \bm{\phi}_i & \bm{\phi}_i \bm{\phi}_i^{\text{T}} + \bm{\Xi}_i
      \end{bmatrix} \bm{\mu}_{p+1} \\
    & \quad \quad \quad - \text{tr} \left( \bm{\Omega}_{p+1} 
      \begin{bmatrix} 
      1 & \bm{\phi}_i^{\text{T}} \\
      \bm{\phi}_i & \bm{\phi}_i \bm{\phi}_i^{\text{T}} + \bm{\Xi}_i
      \end{bmatrix} \right) \bigg].
  \end{align*}
	
	\section*{Session info}

\begin{knitrout}
\definecolor{shadecolor}{rgb}{0.969, 0.969, 0.969}\color{fgcolor}\begin{kframe}
\begin{alltt}
\hlstd{devtools}\hlopt{::}\hlkwd{session_info}\hlstd{()}
\end{alltt}
\begin{verbatim}
## - Session info ---------------------------------------------------------------
##  setting  value                       
##  version  R version 3.6.3 (2020-02-29)
##  os       macOS  10.16                
##  system   x86_64, darwin15.6.0        
##  ui       X11                         
##  language (EN)                        
##  collate  en_US.UTF-8                 
##  ctype    en_US.UTF-8                 
##  tz       Europe/Brussels             
##  date     2021-04-06                  
## 
## - Packages -------------------------------------------------------------------
##  package      * version date       lib source        
##  assertthat     0.2.1   2019-03-21 [1] CRAN (R 3.6.0)
##  backports      1.1.6   2020-04-05 [1] CRAN (R 3.6.2)
##  callr          3.4.3   2020-03-28 [1] CRAN (R 3.6.2)
##  cli            2.0.2   2020-02-28 [1] CRAN (R 3.6.0)
##  colorspace     1.4-1   2019-03-18 [1] CRAN (R 3.6.0)
##  crayon         1.3.4   2017-09-16 [1] CRAN (R 3.6.0)
##  desc           1.2.0   2018-05-01 [1] CRAN (R 3.6.0)
##  devtools       2.3.0   2020-04-10 [1] CRAN (R 3.6.3)
##  digest         0.6.25  2020-02-23 [1] CRAN (R 3.6.0)
##  ellipsis       0.3.0   2019-09-20 [1] CRAN (R 3.6.0)
##  evaluate       0.14    2019-05-28 [1] CRAN (R 3.6.0)
##  fansi          0.4.1   2020-01-08 [1] CRAN (R 3.6.0)
##  formatR        1.7     2019-06-11 [1] CRAN (R 3.6.0)
##  fs             1.4.1   2020-04-04 [1] CRAN (R 3.6.2)
##  glue           1.4.0   2020-04-03 [1] CRAN (R 3.6.2)
##  hms            0.5.3   2020-01-08 [1] CRAN (R 3.6.0)
##  htmltools      0.5.0   2020-06-16 [1] CRAN (R 3.6.2)
##  httr           1.4.1   2019-08-05 [1] CRAN (R 3.6.0)
##  kableExtra     1.1.0   2019-03-16 [1] CRAN (R 3.6.0)
##  knitr        * 1.28    2020-02-06 [1] CRAN (R 3.6.0)
##  lifecycle      0.2.0   2020-03-06 [1] CRAN (R 3.6.0)
##  magrittr       1.5     2014-11-22 [1] CRAN (R 3.6.0)
##  memoise        1.1.0   2017-04-21 [1] CRAN (R 3.6.0)
##  munsell        0.5.0   2018-06-12 [1] CRAN (R 3.6.0)
##  pillar         1.4.3   2019-12-20 [1] CRAN (R 3.6.0)
##  pkgbuild       1.0.6   2019-10-09 [1] CRAN (R 3.6.0)
##  pkgconfig      2.0.3   2019-09-22 [1] CRAN (R 3.6.0)
##  pkgload        1.0.2   2018-10-29 [1] CRAN (R 3.6.0)
##  prettyunits    1.1.1   2020-01-24 [1] CRAN (R 3.6.0)
##  processx       3.4.2   2020-02-09 [1] CRAN (R 3.6.0)
##  ps             1.3.2   2020-02-13 [1] CRAN (R 3.6.0)
##  R6             2.4.1   2019-11-12 [1] CRAN (R 3.6.0)
##  RColorBrewer * 1.1-2   2014-12-07 [1] CRAN (R 3.6.0)
##  Rcpp           1.0.4.6 2020-04-09 [1] CRAN (R 3.6.3)
##  readr          1.3.1   2018-12-21 [1] CRAN (R 3.6.0)
##  remotes        2.1.1   2020-02-15 [1] CRAN (R 3.6.0)
##  rlang          0.4.5   2020-03-01 [1] CRAN (R 3.6.0)
##  rmarkdown      2.1     2020-01-20 [1] CRAN (R 3.6.0)
##  rprojroot      1.3-2   2018-01-03 [1] CRAN (R 3.6.0)
##  rstudioapi     0.11    2020-02-07 [1] CRAN (R 3.6.0)
##  rvest          0.3.5   2019-11-08 [1] CRAN (R 3.6.0)
##  scales         1.1.0   2019-11-18 [1] CRAN (R 3.6.0)
##  sessioninfo    1.1.1   2018-11-05 [1] CRAN (R 3.6.0)
##  stringi        1.4.6   2020-02-17 [1] CRAN (R 3.6.0)
##  stringr        1.4.0   2019-02-10 [1] CRAN (R 3.6.0)
##  testthat       2.3.2   2020-03-02 [1] CRAN (R 3.6.0)
##  tibble         3.0.0   2020-03-30 [1] CRAN (R 3.6.2)
##  usethis        1.6.0   2020-04-09 [1] CRAN (R 3.6.3)
##  vctrs          0.2.4   2020-03-10 [1] CRAN (R 3.6.0)
##  viridisLite    0.3.0   2018-02-01 [1] CRAN (R 3.6.0)
##  webshot        0.5.2   2019-11-22 [1] CRAN (R 3.6.0)
##  withr          2.1.2   2018-03-15 [1] CRAN (R 3.6.0)
##  xfun           0.13    2020-04-13 [1] CRAN (R 3.6.2)
##  xml2           1.3.1   2020-04-09 [1] CRAN (R 3.6.2)
## 
## [1] /Library/Frameworks/R.framework/Versions/3.6/Resources/library
\end{verbatim}
\end{kframe}
\end{knitrout}

\end{document}